%% file: mutation_arxiv.tex
\documentclass[11pt,letterpaper,showkeys,superscriptaddress,sort&compress,rmp,round,floatfix,reprint]{revtex4-1}
\usepackage{snapshot}
\usepackage{amsmath,amsfonts}
\usepackage[T1]{fontenc}
\usepackage{mathptmx,helvet,tikz,pgf,gnuplot-lua-tikz,subfig,paralist}
\usepackage[utf8]{inputenc}
\usepackage{hyperref}
\newcommand{\barz}{\overline{\mathbf{z}}}
\newcommand{\bars}{\overline{s}}
\newcommand{\hats}{\hat{s}}
\newcommand{\hatp}{\hat{p}}
\newcommand{\hatc}{\hat{C}}
\newcommand{\hatk}{\hat{\kappa}}
\newcommand{\barp}{\overline{p}}
\newcommand{\barc}{\overline{C}}
\newcommand{\bark}{\overline{\kappa}}
\newcommand{\dpart}[2]{\frac{\partial #1}{\partial #2}}
\newcommand{\dd}[2]{\frac{d #1}{d #2}}
\newcommand{\Gmatrix}{\mathbf{G}}
\newcommand{\detG}{|\mathbf{G}|}
\newcommand{\detGk}{|\mathbf{G}^{\kappa}|}
\newcommand{\detGkval}{G_{\kappa p}G_{Cp} - G_{C \kappa}G_{p}}
\newcommand{\detGp}{|\mathbf{G}^{p}|}
\newcommand{\detGpval}{G_{\kappa}G_{Cp} - G_{ C \kappa}G_{\kappa p}}
\newcommand{\detGc}{|\mathbf{G}^{C}|}
\newcommand{\detGcval}{G_{\kappa}G_{p} - G_{\kappa p}^{2}}

\newcommand{\barsvalue}{\theta + (G_{\kappa C} + \barc\bark)T }
\newcommand{\uvec}{\mathbf{u}}
\newcommand{\zvec}{\bar{\mathbf{z}}}
\newcommand{\intinfty}{\int_{-\infty}^{\infty}}

\newcommand{\rhock}{\rho_{C \kappa}}
\newcommand{\rhocp}{\rho_{C p}}
\newcommand{\rhokp}{\rho_{\kappa p}}
\newcommand{\barpp}{\barp^{\prime}}
\newcommand{\barppp}{\barp^{\prime\prime}}
\newcommand{\betavec}{\mathbf{\beta}}
\newcommand{\betavecp}{\betavec^{\prime}}
\newcommand{\betavecpp}{\betavec^{\prime\prime}}
\newcommand{\betavechat}{\hat{\betavec}}
\newcommand{\dpdt}{\dd{\barp}{t}}
\newcommand{\dcdt}{\dd{\barc}{t}}
\newcommand{\gvec}{\mathbf{g_\mathrm{p}}}
\newcommand{\piovertwo}{\frac{\pi}{2}}
\DeclareMathOperator{\Cov}{Cov}

\begin{document}

\title{Evolution of female choice and age-dependent male traits with
  paternal germ-line mutation\footnote{Submitted to Ecology and
    Evolution} \footnote{Copyright 2013 Joel J. Adamson.  This is a
    working draft licensed under the Creative Commons
    Attribution-Non-commercial-ShareAlike.  Please contact the author
    for more details}}

\author{Joel James Adamson}
\affiliation{Ecology, Evolution and Organismic Biology\\
  CB \#3280 \\
  University of North Carolina, Chapel Hill, NC 27599\\
  919-843-2320} \email{adamsonj@ninthfloor.org}

\bibliographystyle{chicago}
\begin{abstract}
  Several studies question the adaptive value of female preferences
  for older males.  Theory and evidence show that older males carry
  more deleterious mutations in their sperm than younger males carry.
  These mutations are not visible to females choosing mates.
  Germ-line mutations could oppose preferences for ``good genes.''
  Choosy females run the risk that offspring of older males will be no
  more attractive or healthy than offspring of younger males.
  Germ-line mutations could pose a particular problem when females can
  only judge male trait size, rather than assessing age directly.  I
  ask whether or not females will prefer extreme traits, despite
  reduced offspring survival due to age-dependent mutation.
  I use a quantitative genetic model to examine the evolution of
  female preferences, an age-dependent male trait, and overall health
  (``condition'').  My dynamical equation includes mutation bias that
  depends on the generation time of the population.  I focus on the
  case where females form preferences for older males because male
  trait size depends on male age.
  My findings agree with good genes theory.  Females at equilibrium
  always select above-average males.  The trait size preferred by
  females directly correlates with the direct costs of the preference.
  Direct costs can accentuate the equilibrium preference at a higher
  rate than mutational parameters.  Females can always offset direct
  costs by mating with older, more ornamented males.
  Age-dependent mutation in condition maintains genetic variation in
  condition and thereby maintains the selective value of female
  preferences.  Rather than eliminating female preferences, germ-line
  mutations provide an essential ingredient in sexual selection.
\end{abstract}

\keywords{sexual selection; mate choice; life history; mutation; costly preferences}
\maketitle

\section{Introduction}
\label{sec:intro}


Sexual selection theory yields predictions for evolution of
extravagant signals in males and preferences in females.  Researchers
categorize signals into two categories:
\begin{inparaenum}[(1)]
\item traits that signal direct benefits, i.e.\ choosy females
  themselves gain higher fitness; or
\item traits that signal indirect (genetic) benefits to females.
  Choosy females incur viability or fecundity costs during mate
  choice, but matings yield higher offspring fitness
  \citep{jones_colloquium_2009,andersson_sexual_1994}.  Costs to
  females may occur in the form of exposure to predators, pathogens
  and parasites, or reduced chances of mating.
\end{inparaenum}
Research on indirect benefits often focuses on costly preferences for
traits that signal heritable condition of trait-bearing males
\citep[``condition-dependent'' traits;][]{andersson_evolution_1986}.
As the name suggests, the size of condition-dependent traits depends
directly on the health of the male carrying the trait.  Thus females
can produce healthy offspring by exercising preferences for extreme
values of condition-dependent traits.  If selection were to act
unchecked by mutation, genetic variation in condition of males would
disappear, since the healthiest males survive best and gain most
matings.  A female in this situation optimally mates randomly, since
all males carry the same genes for overall health.  Selection would
act against females expressing costly choices, since they gain no net
benefits \citep[the ``lek
paradox'';][]{rowe_lek_1996,kirkpatrick_evolution_1991}.  Selection
maintaining female preferences therefore requires biased mutation in
condition, i.e.\ repeated appearance of deleterious mutations in male
offspring \citep{iwasa_evolution_1991}.


Another line of research focusing on male traits suggests that males
benefit from age-dependent investment in their signals
\citep{kokko_good_1998,kokko_evolutionarily_1997}.
\citet{proulx_older_2002} showed that a male's optimal life-history
strategy depends on his condition: high-condition males will conserve
resources and signal less at young ages, then increase their signaling
into old age.  Lower-condition males optimally signal at a high rate
when they are young, since they have less chance of surviving to old
age.  These studies predict that females evolve preferences for
older-aged males.  Theory and empirical work throughout the history of
research on sexual selection support the hypothesis that selection
favors preferences for older- or middle-aged males
\citep{brooks_can_2001}.  I previously showed that reduced adult
mortality and age-dependent signals promote sexual selection by
reducing selection against costly traits \citep{adamson_peerj_2013}.
When male traits start out small, selection cannot eliminate traits
that grow large later in life.  Long-term data sets in mammals support
this hypothesis by showing increased heritability and greater marginal
benefits of sexually selected traits later in life
\citep{poissant_quantitative_2008,pemberton_mating_2004,courtiol_natural_2012}.


Reduced adult mortality and age-dependent expression thus favor
preferences for older males.  This prediction seems to provide a
convenient explanation for older-male preferences.  However,
\citet{hansen_good_1995} raised four objections to the feasibility of
selection favoring older-male preferences.  Among these objections was
the substantial number of deleterious mutations in the paternal
germ-line
\citep{kong_rate_2012,sayres_variations_2011,ellegren_characteristics_2007,goetting-minesky_mammalian_2006,moller_sexual_2003,bartosch-harlid_life_2003,radwan_male_2003,hansen_age-and_1999}.
Spermatogonia, the stem cells that mature into spermatocytes, undergo
frequent mitotic divisions during the reproductive life of a male
animal.  Mutations in sperm cells increase with male age, even when
mutation rate remains constant over the lifespan.  These germ-line
mutations negatively affect the heritability of male condition.  Males
who survive into attractive old age could yield offspring who do not
survive well \citep[see][who showed this could be true even without
germ-line mutations]{hansen_good_1995}.  Germ-line mutations could
therefore negatively impact the evolution of costly female
preferences.  Maintenance of preferences depends on the genetic
correlation between preferences and condition
\citep{Mead2004264,iwasa_evolution_1991}.  If females with strong
preferences give rise to mutation-ridden offspring with lowered
condition, preferences could correlate weakly or even negatively with
condition.



Here I ask whether or not selection will maintain costly female
preferences for older males when
\begin{inparaenum}[(1)]
\item females can only assess trait size; and
\item a male's mutational contribution depends on his age.
\end{inparaenum}
I focus on cases where trait size explicitly depends on age.  Will a
population under sexual selection achieve equilibria with females
expressing directional sexual preferences, rather than favoring
average or optimal males?  When age-dependent traits and age-dependent
mutation occur together, females preferring above-average traits could
select males yielding less fit, mutant offspring.  Natural selection
could therefore eliminate costly female preferences.  I will show,
using a quantitative genetic model of preferences, traits and
condition, that the process supplying necessary mutation bias
(germ-line mutation) coincides with the process facilitating sexual
selection (age-dependent traits).  Selection maintains female
preference for male traits in this model, including cases with
preferences for older males.  Age-dependent paternal mutation
reinforces the evolution of costly female preferences, rather than
hindering it.

\section{Model}
\label{sec:model}

\subsection{Phenotypic and population model}
\label{sec:phenotypes}

I will explore the above questions using the evolutionary dynamics of
mean phenotypes \citep{lande_quantitative_1982}.  Individuals in the
large population express a phenotype composed of a set of characters
with non-overlapping genetic components (i.e.\ no pleiotropy).  I
assume that environmental variance has mean $0$ and therefore ignore
it when considering the evolution of mean phenotypes.  The above
considerations imply the equality of genetic and phenotypic variances,
covariances and correlations \citep{iwasa_evolution_1991}.
Quantitative genetic models typically ignore the effects of epistasis
and dominance for simplicity, as I do here.  I wish to isolate the
effects of selection and mutation and therefore ignore the potential
effects of drift and migration.

The phenotype consists of three traits:
\begin{inparaenum}[(1)]
\item a growth parameter $\kappa$,
\item a female preference $p$ and
\item intrinsic viability or ``condition'' $C$.
\end{inparaenum}
\citet{iwasa_evolution_1991} used a similar model of
condition-dependent traits.  Males of condition $C$ and growth
parameter $\kappa$ express the signaling trait $s$ as a function of
age ($x$):
\begin{equation}
  \label{eq:trait}
  s (x)  = \theta + C \kappa x
\end{equation}
where $x$ represents age and $\theta$ represents the optimal trait
value for viability selection.  Note that all juvenile males produce
the optimal trait size.

Only females express the preference $p$ and females display the trait
at the optimum $\theta$.  Females express a preference independent of
age and condition.  A unimodal function expresses the relative
frequency of matings of males with phenotype $s$ to females of
phenotype $p$:
\begin{equation}
  \label{eq:unimodal}
  \phi (s|p) = \exp \left(-\frac{(s-p)^{2}}{2\sigma^{2}}\right)
\end{equation}
where $\sigma$ represents the width (``standard deviation'') of the
preference function \citep{lande_models_1981}.  Smaller values of
$\sigma$ indicate preferences more tightly concentrated around $p$.
Females have the highest relative frequency of mating with males whose
size $s$ matches their preferred value $p$.

Phenotypic condition ($C$) remains constant throughout an individual's
life, but an individual male's germ-line genotypic value (breeding
value) of $C$ decays according to the linear function
\begin{equation}
  \label{eq:mutation}
  \mu (x) = -\mu x
\end{equation}
where $\mu$ represents a constant, phenotypically standardized
mutational effect (e.g. one could measure $\mu$ in units of $\theta$).
A male of age $x$ on average delivers genetic value $\barc - \mu x$ to
his offspring.

I assume that the population grows exponentially according to
\begin{equation}
  \label{eq:exponential}
  \dd{N}{t} = rN 
\end{equation}
where $r$ represents the largest real root in $r$ of
\begin{equation}
  \label{eq:euler}
  \int_{0}^{ \omega} e^{-rx}l (x)m (x) dx = 1.
\end{equation}
The hazard function $l (x)$ represents survival to age $x$, $m (x)$
represents fecundity at age $x$ and $\omega$ symbolizes the oldest age
in the population.  I assume these vital rates remain roughly constant
over short periods of little phenotypic change \citep[as
in][]{lande_quantitative_1982}.  I use \emph{generation time} 
\begin{equation}
  \label{eq:T}
  T = \int_{0}^{\omega}xe^{-rx}l (x)m (x) dx
\end{equation}
as a parameter expressing the basic structure of the life-history.
Readers can also conceptualize $T$ as the average age of breeding
adults.  Small values of $T$ represent populations of individuals with
relatively short lives, whereas larger values of $T$ represent
populations of individuals with longer lifespans.
Equation~(\ref{eq:euler}) holds at stable age distribution, hence I
assume weak selection.  By weak selection, I mean that any disturbance
of stable age distribution caused by change in the mean phenotype will
quickly converge before more phenotypic change occurs.  For example,
if mean condition increases, survival to old age could also increase,
taking the population temporarily out of stable age distribution.
Then I assume that the age distribution will restabilize within a few
generations, before mean condition changes again and brings about the
next demographic disturbance.  Therefore I can assume the population
remains in stable age distribution over the course of evolutionary
change \citep[see
][]{charlesworth_natural_1993,lande_quantitative_1982}.

I use the dynamical equation
\begin{equation}
  \label{eq:gbeta}
  \dd{\zvec}{t} = \left(\frac{1}{2}\right)\left(\Gmatrix \beta - T\uvec\right)
\end{equation}
where $\beta$ represents the selection gradient in
Equation~(\ref{eq:beta}), $\Gmatrix$ contains variance $G_{i}$ for
trait $i$ on the diagonal and covariances $G_{ij}$ for traits $i,j$
off-diagonal.  The expression $T\uvec$ represents the mutational input
into the next generation, where the vector $\uvec$ contains mutation
rates in each trait.  For current purposes
\begin{equation}
  \label{eq:mu}
  \uvec =
  \begin{pmatrix}
    0 \\
    0 \\
    \mu
  \end{pmatrix}
\end{equation}
where $\mu$ signifies a constant per-unit-time probability of mutation
in spermatogenesis (see Equation~(\ref{eq:mutation})).  
Equation~(\ref{eq:gbeta}) carries a factor of $\frac{1}{2}$ due to
sex-limited expression.  Equation~(\ref{eq:gbeta}) derives from
\citep{lande_quantitative_1982} with the addition of sex-limited
expression and the mutation term.  \citet{charlesworth_natural_1993},
\citet{iwasa_evolution_1991} and others have derived similar equations
for both continuosuly differentiable time and discrete-time models
\citep[see also][]{kokko_unifying_2006}.

\subsection{ Fitness }
\label{sec:demography-fitness}

Viability selection on the trait $s$ at age $x$ follows the Gaussian
function
\begin{equation}
  \label{eq:mfit}
  w(s,C) = \exp \left(-\frac{(s (x) - \theta)^{2}}{2C^{2}}\right)
\end{equation}
such that males carrying $s$ smaller and larger than $\theta$ suffer
viability costs, dependent on the condition of the bearer.  The
fitness of males with higher $C$ will slope off more gradually than
that of males with lower $C$ \citep{kotiaho_costs_2001}.  I calculate
total male fitness by multiplying viability by mating success.  The
mating system follows a polygynous model.  Males gain mating success
\begin{equation}
  \label{eq:U}
  U (s) = \intinfty f (p) \phi (s|p) dp
\end{equation}
given female population $f (p)$.  Substituting
Equation~(\ref{eq:trait}) into Equation~(\ref{eq:mfit}), males gain a
total fitness
\begin{subequations}
  \label{eq:fitness}
  \begin{equation}
    \label{eq:matingsuccess}
    W_{m} = \exp \left(- \frac{(\kappa x)^{2}}{2}\right)
    \intinfty f (p) \phi (s|p) dp .
  \end{equation}
  during any short time interval $\Delta t$.

  Viability selection on females depends on condition, preference
  value, and the average male signaling trait:
  \begin{equation}
    \label{eq:wf}
    W_{f} = \exp\left(-\frac{b(p-\bars)^{2}}{2C^{2}}\right).
  \end{equation}
\end{subequations}
I assume that female fecundity varies independently of age and
condition, and that all females (across the distribution of $C$) gain
the same average mating success during any time interval $\Delta t$.
Female fecundity as a function of condition could seem more realistic,
but my goal is to isolate the effects of costly female preferences.  I
therefore assume that female fitness varies only due to viability
costs generated by mate choice.

I calculate selection gradients by differentiating the natural
logarithm of total fitness of males and females with respect to
specific phenotypic traits and evaluating these partial derivatives at
the population mean for the trait, with age set to the generation
time:
\begin{subequations}
  \label{eq:gradients}
  \begin{align}
    \dpart{\log W_{m}}{\kappa} = & -\kappa x^{2} -
    \frac{Cx}{\sigma^{2}} \left(\frac{\intinfty (s-p) f (p) \phi
        (s|p)dp}
      {\intinfty f (p) \phi (s|p)dp}\right) \label{eq:kappagrad}\\
    \dpart{\log W_{f}}{p} = & - \frac{b (p-\bars)}{C^{2}} \\
    \dpart{\log W_{m}}{C} = & \frac{1}{C^{3}}\left( \frac{\intinfty
        (s-p)^{2} f
        (p) \phi (s|p)dp} {\intinfty f (p) \phi (s|p)dp} \right)\label{eq:Cmgrad}\\
    \dpart{\log W_{f}}{C} = & \frac{b (p-\bars)}{C^{3}}.
  \end{align}
\end{subequations}
The denominators in Equations~(\ref{eq:kappagrad}) and
~(\ref{eq:Cmgrad}) result from logarithmic differentiation of
Equation~(\ref{eq:matingsuccess}).

An important step in the interpretation of the results involves
assessing female ability to quantity underlying traits, particularly
condition ($C$).  Females in this model can only use the trait ($s$)
itself to do this.  I will assume that a female's ability to assess
underlying condition corresponds to the correlation of the underlying
trait to the signal.  For some function $h:\mathbb{R}^{n}\rightarrow
\mathbb{R}$ I calculate the covariance of $h$ and some component
$y_{i}$ of its domain as
\begin{equation}
  \label{eq:func-cov-general}
  G_{f{y_{i}}} = \dpart{h}{y_{i}}\vert_{\bar{y_{i}}} G_{y_{i}}.
\end{equation}
Accordingly I obtain the covariances between the trait ($s$) and
condition and growth:
\begin{subequations}
  \label{eq:s-corr}
  \begin{align}
    G_{s\kappa} = & \barc T G_{\kappa} \label{eq:rho-sk} \\
    G_{sC} = & \bark T G_{C}\label{eq:rho-sc}.
  \end{align}
\end{subequations}
Growth and trait always positively covary.  The males with the largest
traits in the population will tend to have the largest growth
coefficients.  Equation~(\ref{eq:rho-sc}) can become negative only if
$\bark < 0$.  Increasing generation time will increase the absolute
value of these covariances in both cases.  This accords with the
higher variance in trait sizes afforded by longer generation time.
Longer generation time means more old males in the population, hence
more large traits and larger variances.
\section{Results}
\label{sec:results}

\subsection{Selection gradient}
\label{sec:gradient}
The selection gradient illustrates the basic direction and magnitude
of evolutionary change in the three phenotypic characters.  The
largest component of character change comes from the multiplication of
the selection gradient for each character ($\beta_{i}$) with its
variance ($G_{i}$).  I refer to all other effects in any equation as
``indirect.''  Therefore looking directly at the selection gradient
shows the basic structure of the model:
\begin{subequations}
  \label{eq:beta}
  \begin{align}
    \beta_{\kappa} = \dpart{\log{W_{m}}}{\kappa} \vert_{\barz,T}
    = &  T \left(\frac{\barc (\barp-\bars)}{\sigma^{2}} - \bark T\right) \label{eq:beta-male}\\
    \beta_{p} = \dpart{\log{W_{f}}}{ p} \vert_{\barz,T} =
    & - \frac{b (\barp-\bars)}{\barc^{2}}\label{eq:beta-female}\\
    \beta_{C} = \dpart{\log{W_{m}}}{ C} \vert_{\barz,T} +
    \dpart{\log{W_{f}}}{ C} \vert_{\barz,T}= & (1+b) \left(\frac {(\barp-\bars)^{2}}{\barc^{3}}\right)\label{eq:beta-condition}
  \end{align}
\end{subequations}
Equation~(\ref{eq:beta-male}) shows that
\begin{inparaenum}[(1)]
\item selection on males intensifies as generation time ($T$)
  increases;
\item the life-history accentuates splitting selection on males into
  sexual (the first term in parentheses) and viability components (the
  second term in parentheses);
\item males carrying different somatic values of condition ($C$) or
  growth ($\kappa$) can display similar trait values with varying
  linearized fitness in the $\kappa$-dimension.
\end{inparaenum}
Direct selection on female preference
(Equation~(\ref{eq:beta-female})) leads to negative character change
unless the preferred size ($p$) falls below the average trait
($\bars$).  Intensity of selection against choice attenuates with
increasing condition.  Selection on condition
(Equation~(\ref{eq:beta-condition})) remains strictly positive at all
trait values and intensifies with the direct costs of choice.  Only
selection on trait growth ($\kappa$) depends on generation time ($T$).
More intense (sexual and viability) selection occurs on male traits in
populations with longer-lived life-histories.  The selection gradient
here predicts lower values of $\kappa$ at longer generation times.
This follows from the model of trait growth, as populations with
longer generation times will have more old males with larger traits.

\subsection{Equilibria}
\label{sec:equilibria}

I find equilibria in $\bar{\mathbf{z}}$ by setting the left-hand side
of Equation~(\ref{eq:gbeta}) equal to $0$ and using Cramer's Rule,
substituting the values in Equation~(\ref{eq:beta}) for $\betavec$:
\begin{subequations}
  \label{eq:eq}
  \begin{align}
    \frac{\barc (\barp-\bars)}{\sigma^{2}} - \bark T = & \frac{\mu \detGk}{\detG} \\
    \frac{b (\barp-\bars)}{\barc^{2}} =  & \frac{T\mu \detGp}{\detG}\\
    (1+b) \left(\frac {(\barp-\bars)^{2}}{\barc^{3}}\right) = &
    \frac{T\mu \detGc}{\detG}
  \end{align}
\end{subequations}
where
\begin{subequations}
  \label{eq:Gvals}
  \begin{align}
    \detGk = & \detGkval\\
    \detGp = & \detGpval \\
    \detGc = & \detGcval\label{eq:detGc}
  \end{align}
\end{subequations}
form the principal minors of the upper two rows of $\Gmatrix$.
Readers can find $|\Gmatrix^{i}|$ by replacing column $i$ of
$\Gmatrix$ with $T\uvec$ and taking the determinant of the resulting
matrix.  Equation~(\ref{eq:detGc}) measures the basic intensity of
direct correlation between male growth ($\kappa$) and female
preference ($p$).  This expression simplifies to $\detGc = G_{k}G_{p}
(1 - \rhokp^{2})$ and remains non-negative over all values of
$\rhokp$.
\begin{equation}
  \label{eq:shat}
  \hat{s} = \barsvalue
\end{equation}
describes the equilibrium value of $\bars$, found by substituting from
Equations~(\ref{eq:eq}) and averaging over the male population, using
the approximation $E[C\kappa x] \approx \overline{C\kappa} T$.  I have
assumed here that $\Cov (C\kappa,x) \approx 0$ due to the slow rate of
phenotypic change.




Assuming that $b > 0$ I then solve Equations~(\ref{eq:eq}) for
equilibria in all three variables, yielding
\begin{subequations}
  \begin{align}
    \hatk = & \frac{b^{5}\detG^{3}\detGc^{2}}{\sigma^{2} (1+b)^{3}T^{3}\mu^{2}\detGp^{5}} -
    \frac{\mu \detGk}{T\detG} \\
    \hatp = & \frac{b^{3}\detG\detGc^{2}}{(1+b)^{2}T\mu\detGp^{3}} + \hats \label{eq:p-eq}\\
    \hatc = & \frac{b^{2}\detG\detGc}{(1+b)T\mu \detGp^{2}}\label{eq:c-eq}
  \end{align}
\end{subequations}

The equilibrium in male trait growth ($\kappa$) includes a positive
component (as long as $\detG$ remains positive) and a negative
component, mirroring the mating and viability components of the
selection coefficient (Equation~(\ref{eq:beta-male})).  The negative
component increases with mutation size ($\mu$) and decreases with
generation time.  Here the two forces oppose each other, whereas in
the other equilibria they multiply together.

The equilibrium in female preference ($p$) lies above the equilibrium
value of the male trait ($\bars$) as long as $\detGp > 0$, easily seen
in this alternate representation:
\begin{equation}
  \label{eq:alternate-phat}
  \hatp - \hats = \frac{\hatc^{2} T\mu\detGp}{b}.
\end{equation}
The difference between the equilibrium preference and the equilibrium
trait will shrink as mutation size ($\mu$) and generation time ($T$)
increase, or as the scaling parameter of costs ($b$) gets smaller.

The equilibrium given by Equation~(\ref{eq:c-eq}) shows that
mutation-selection balance in $C$ remains positive and diminishes as
mutation rate increases.  The mutation-selection balance value for
$\barc$ will increase as the scaling parameter of female viability
costs ($b$) increases, or as the correlation ($\rhokp$) between
preference ($p$) and growth ($\kappa$) decreases.

\subsection{Interpretation}
\label{sec:interp}

I will interpret the model in terms of what I call basic good genes
theory (``good genes'').  Selection will favor female preferences that
improve offspring condition relative to the offspring condition of
females with other preferences
\citep{kokko_fisherian_2001,jones_mate_2009}.  I will phrase the
interpretations in terms of the position of mean female preference
relative to the mean male trait, and the effects of life-history
parameters (e.g.\ increasing generation time).

\begin{figure*}[t!]
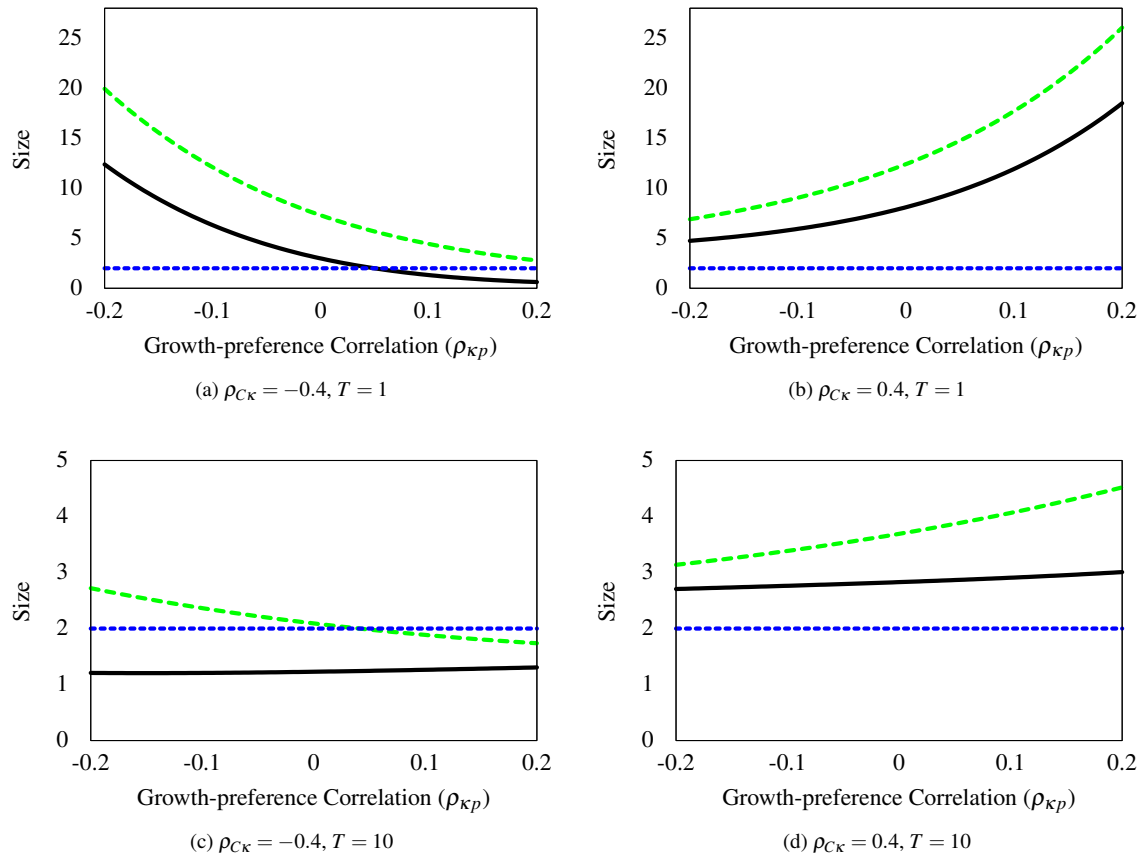

    \begin{center}
      \subfloat[$\rhock = -0.4$, $T = 1$]{\label{subfig:corr1} \input
        r_kp_01.tex } \subfloat[$\rhock = 0.4$, $T =
      1$]{\label{subfig:corr2} \input r_kp_02.tex
      }\\
      \subfloat[$\rhock = -0.4$, $T = 10$]{\label{subfig:corr3} \input
        r_kp_03.tex } \subfloat[$\rhock = 0.4$, $T =
      10$]{\label{subfig:corr4} \input r_kp_04.tex }
    \end{center}
    \caption{The equilibrium values of trait (solid black) and
      preference (dashed green) as a function of the genetic
      correlation between male growth and female preference ($\rhocp =
      0.2$).  Panels show positive (right-hand column) and negative
      (left-hand column) values of $\rhock$, at shorter (top row) and
      longer life-histories (bottom row).  The blue dashed line
      indicates the optimal male trait size $\theta$.  Mutational
      effect size ($\mu$) equals the optimal size of the trait
      ($\theta$) in all panels.}
    \label{fig:size_corr}
\end{figure*}

\begin{figure*}[!]
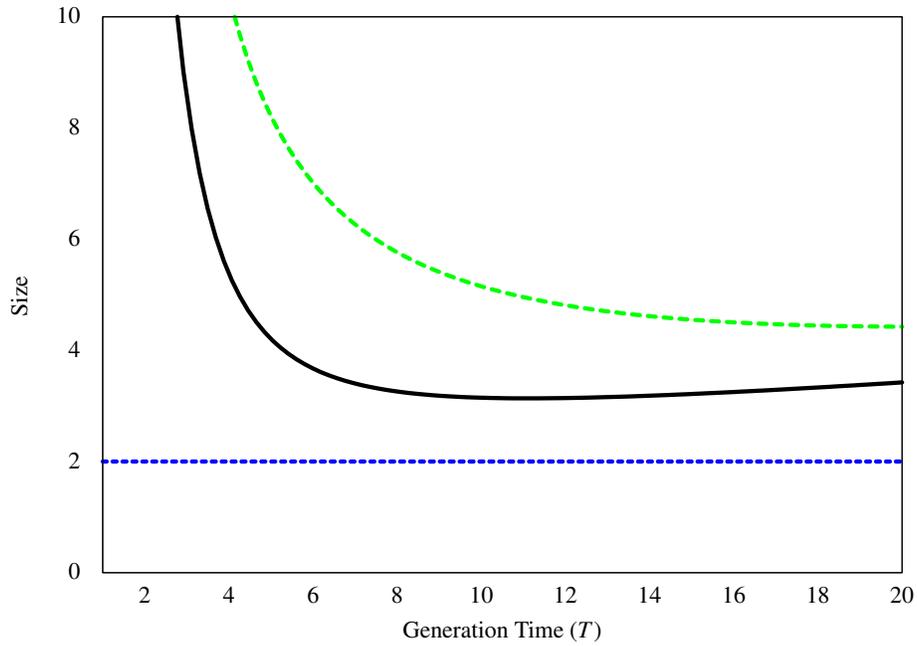
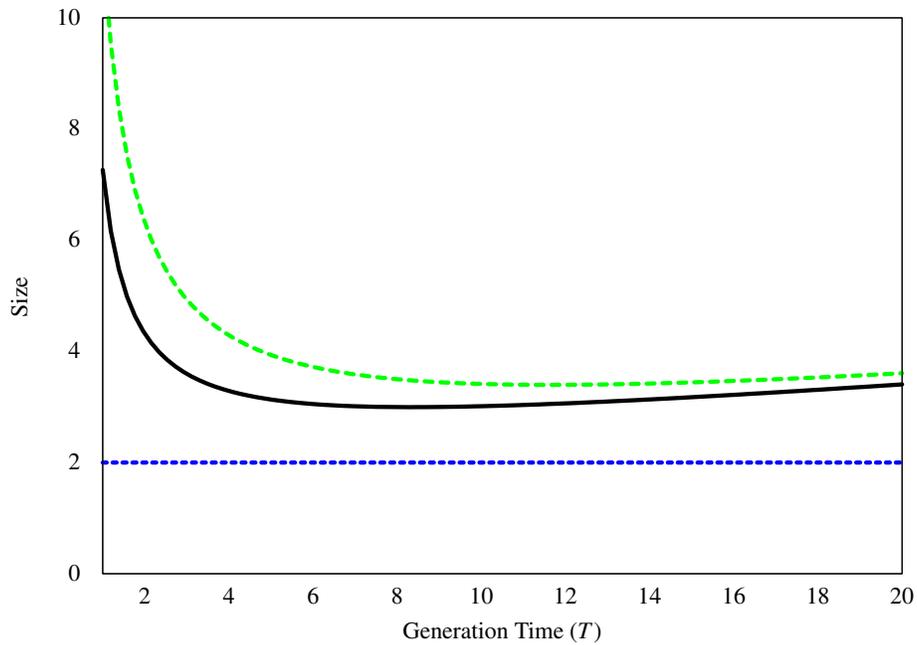

    \begin{center}
      \subfloat[$\rhock = 0.4$, $\mu = \theta$]{\label{fig:munotthatbig}\input tpeqb01_pos.tex }\\
      \subfloat[$\rhock = 0.4$, $\mu = 5\theta$]{\label{fig:mubig}\input
        tpeqb01_neg.tex }

      \caption{Equilibrium trait (solid black) and preference (dashed
        green) as a function of generation time, with varying mutational
        effect sizes ($\mu$).  The blue dashed line indicates the optimal
        male trait size $\theta$.  Comparison of the two panels shows that
        increasing mutational effect size decreases the difference between
        equilibrium trait and equilibrium preference.}
      \label{fig:mu-effect}
    \end{center}
  
\end{figure*}

The three genetic correlations from $\Gmatrix$
(Figure~\ref{fig:size_corr}) affect the hypothetical position of
female preference relative to the male trait.  The terms of
Equation~(\ref{eq:Gvals}) translate into correlations using the
relation $\rho_{ij} = \frac{G_{ij}}{\sqrt{G_{i}G_{j}}}$, which forms
an upper bound for genetic covariances.  I assume throughout this
analysis that $\rhocp > 0$ since this condition forms a prerequisite
for the evolution of preferences for indicator traits.  I especially
focus on cases where selection favors females that choose males with
small growth parameters ($\kappa$), since trait size ($s$) does not
reveal either condition ($C$) or growth rate ($\kappa$) directly.
This situation occurs when males with high growth parameters tend to
have low-condition offspring, i.e. $\rhock < 0$.  Negative
correlations might evolve under trade-offs between trait size and
viability.  Readers should keep in mind that
Figure~\ref{fig:size_corr} and Figure~\ref{fig:mu-effect} were created
using a fairly large mutational effects.  For example, for a
mutational effect of $\mu = \theta$ (Figure~\ref{fig:size_corr} and
Figure~\ref{fig:munotthatbig}) the difference in trait size between a
father of age $x$ and his son at that same age equals $-\theta \kappa
x$.  Differences attributable to mutation are even larger in
Figure~\ref{fig:mubig}, where with short to intermediate generation
times we still see females with fairly strong preferences.

Good genes predicts a non-negative difference between equilibrium
female preference and equilibrium male trait.  I base this
interpretation on \citet{iwasa_evolution_1991}, who found equilibrium
female preference proportional to mutational input and inversely
proportional to the scaling parameter of costs ($b$ here and in
\citeauthor{iwasa_evolution_1991}).  Positive mutational bias enables
an honest signal by depressing average male condition (i.e.\
increasing $G_{C}$ on the right-hand side of
Equation~(\ref{eq:rho-sc})).  Females with positive preferences ($p >
\bars$) will produce higher quality offspring than females with
average preferences ($p = \bars$, i.e.\ random mating).  My model
differs from theirs in that female viability costs are proportional to
the difference $\barp - \bars$, and so I evaluate the predictions of
the theory slightly differently.

When growth and condition positively correlate ($\rhock > 0$) the
equilibrium $\kappa$ will lead to average traits above $\theta$, due
to positively reinforcing viability and mating advantages.  Selection
will favor females whose preferences lead to larger $\kappa$ values
(since $\kappa$ and $C$ positively correlate).  The predicted
equilibrium $p$ increases with $\rhokp$.  Good genes predicts
balancing selection at negative values of $\rhokp$, since females with
extraordinarily low preferences will tend to choose small-$\kappa$
males as mates (see Equation~(\ref{eq:rho-sk})), producing
low-condition offspring.

When growth and condition negatively correlate ($\rhock < 0$) the
equilibrium $\kappa$ will lead to small traits, with average trait
smaller than $\theta$ under some values of $\rhokp$.  Selection will
favor females whose preferences lead to smaller $\kappa$ values, since
low-$\kappa$ offspring will have higher $C$.  The predicted
equilibrium $p$ decreases with $\rhokp$.  Again good genes predicts
balancing selection as $\rhokp$ becomes more strongly negative, since
females with especially large preferences will tend to choose
large-traited males as mates, leading to offspring with low condition
(see Equation~(\ref{eq:rho-sc})).  Longer generation time means more
old males will be available for mating, and females could gain good
genes by mating with even older males.  Therefore good genes
universally predicts an increasing divergence between preference and
trait values.  Figure~\ref{fig:size_corr}, on the other hand, shows
that increasing generation time reduces the equilibrium deviation of
preference from average trait, rather than increasing it (see
Equation~(\ref{eq:p-eq})).

%

\section{Discussion}
\label{sec:discussion}

The evolution of age-dependent signals appears to follow the basic
guidelines set down by good genes theory.  Age-dependent sexual
signals could facilitate sexual selection by reserving the production
of costly traits until  older ages.  Life-history strategy theory
shows that honest signaling favors delaying development of costly
signals, and favors preferences for older males.  Older-male
preferences, however, may come with costs arising from the higher
mutation load of older males' sperm.  I have shown that viability
selection does not eliminate sexual selection.  Female preferences can
remain positive despite considerable mutational effects.  Furthermore,
my results show that
\begin{inparaenum}[(1)]
\item sexual selection intensifies as generation time increases and
\item selection on condition intensifies as direct costs of choice
  increase.  
\end{inparaenum}

The results of the model coincide with existing interpretations of
good genes theory, but also contains information about the action of
life-histories.  Basic interpretations of good genes do not include
any effects of age-dependent mutation.  My interpretation of good
genes places limits on the equilibrium deviation of preference from
trait, but does not produce accurate predictions regarding the action
of generation time.  Increasing generation time reduces the departure
of the preference from the trait, rather than increasing it (see
Equation~(\ref{eq:p-eq}) and Figure~\ref{fig:size_corr}).

The failure of the basic theory leads me to consider two additional
hypotheses.  What I call mutational effects balancing selection
(``mutational effects'') predicts that along with the predictions of
basic good genes, the effects of mutation place limits on the adaptive
value of extreme preferences.  This theory roughly corresponds to the
objections to good genes raised by \citet{hansen_good_1995}. Direct
costs balancing selection (``direct costs'') includes some of the
predictions of both prior theories, but predicts that direct costs to
preferences set the limit on the deviation of preference from average
trait.

The mutational effects theory of balancing selection emphasizes that
although selection favors female preferences, mutation places limits
on the adaptive value of preferring older males.  At some point the
marginal benefit in offspring condition will maximize owing to the
higher mutation rate of older males.  The basic predictions of good
genes theory still hold.  Female preferences still lie above male
trait sizes.  The difference between the hypotheses lies in more
accurate predictions based on increasing mutation rate (higher
$T\mu$).  Mutation rate augmentation will decrease the deviation
between trait and preference.  This prediction holds true in the model
(see Equations~(\ref{eq:p-eq}) and
Figure~\ref{fig:size_corr}). Mutational theory also predicts reduction
in the absolute size of the equilibria (Figure~\ref{fig:mu-effect}).
More old males in the population (larger $T$) leads to larger
mutational input, meaning a smaller average benefit.  Correspondingly,
at short generation time, few males live long enough to accumulate
many mutations.  Selection will favor females selecting the oldest
(i.e.\ largest) males they can find.  

Direct costs balancing selection forms a third possibility.
Considering direct costs of female choice makes the prediction of
upper limits on preference more precise.  Direct costs can account for
the observation that as the scaling coefficient of mate choice costs
($b$) increases, the equilibrium preference increases (see
Equation~(\ref{eq:p-eq})).  As choice becomes more directly costly,
selection favors females that secure better genes for their offspring.
Despite the risk of greater mutational input from larger-trait males,
augmenting costs ($b$) creates stronger selection to find good genes.
The fitness differential in offspring condition will offset any
mutational losses in condition \emph{and} losses in female viability.
Readers can verify this in Equation~(\ref{eq:c-eq}) by noting that
equilibrium condition ($\barc$) also increases with $b$.  Although
some results of my model are consistent with the effects predicted by
a theory of mutational effects, direct costs to females form the best
explanation for my results.


%
I can also consider the effectiveness of the age-as-indicator model
\citep{brooks_can_2001}.  Although age does not constitute an
indicator in this model, age does influence the reliability of the
indicator trait (Equation~(\ref{eq:rho-sc})).  When growth and
condition positively correlate, older males will tend to have larger
traits.  Selection will also favor females that prefer larger traits.
Good genes predicts that females will choose older males in all cases.
Direct costs theory adds to the precision of this prediction.  Despite
mutation pressure, older males signal condition in a more reliable
manner.  On the other hand, in some cases when growth and condition
negatively correlate, selection still favors females that choose the
largest traits (Figures~\ref{subfig:corr1} and \ref{subfig:corr3}).
We can interpret this as selection against older males, since older
males will show smaller traits, and the equilibrium $\barp$ still lies
above $\hats$.

Age-dependence and paternal mutation facilitate and maintain selection
for female preferences.  First, increasing life-span facilitates
sexual selection by reducing the power of selection against the trait
\citep{adamson_peerj_2013}.  Secondly, increasing lifespan
increases the mutational input into condition, maintaining genetic
variance and selection for female preferences.  As selection weakens
over the lifespan, traits can become more accentuated without
impacting fitness as strongly.  Mate choice for older males also
contributes to the life-stage separation enabled by the weakening of
selection.  As females produce a broader range of condition in
offspring, viability selection has more variance to work with in the
early stages of life.  Increased lifespan and age-dependence with mate
choice therefore introduce a negative feedback in terms of genetic
variation, viability selection and mate choice.  The lek paradox
disappears as the process enabling trait exaggeration simultaneously
introduces genetic variation into the next generation.

Generalizing my results requires some caution.  If viability selection
weakens with age, enabling more extravagant traits, I expect that
selection will also allow greater mutation accumulation in the trait
itself and loss of variation through genetic drift.  My model makes
several assumptions that do not cover this possibility.  I completely
neglect drift, as well as the genetic mechanisms of dominance,
epistasis and pleiotropy.  Traits and preferences can correlate much
more tightly than I've supposed here, including pleiotropy
\citep{grace_coevolution_2011}.  I chose particular functions for
mathematical convenience that limit the scope of the application.
Furthermore, since I seek the elucidation of theory, I have neglected
details of male and female mating behavior.

I began with the question of whether selection can maintain costly
preferences in the face of mutation pressure.  Age-dependent mutation
appears to supply the necessary genetic variation for sexual selection
to continue.  Contrary to some expectations, direct costs to females
produce the greatest selective incentive for females to express
preferences for extreme male traits.  Direct costs and the continued
input of variation in indirect benefits interact to reinforce sexual
selection on an indicator trait.  Selection maintains costly
preferences supported by, rather than despite, age-dependent mutation.
The question of whether or not age-dependent preferences display the
same patterns remains open.

\begin{acknowledgments}
\label{sec:ack}

I would like to thank Maria Servedio and Haven Wiley for overseeing
this manuscript and giving me invaluable advice on language.  Sumit
Dhole, Caitlin Stern, Courtney Fitzpatrick and Justin Yeh repeatedly
provided input on behavior of actual animals.  I would especially like
to thank the owners of the Scratch Bakery in Durham, NC where the
mathematics of this research really came together while I feasted on
their infamous donut cupcakes.  This research was funded by NSF
DEB-0614166 and NSF DEB-0919018 to Maria Servedio.
\end{acknowledgments}
\bibliography{mutation}

\appendix*
\section{Stability }
\label{sec:appendix}

\begin{figure}[!]
  
    \begin{center}
      \begin{tikzpicture}[thick,domain=0:3,>=stealth]

        \draw[->] (-0.2,0) -- (4.2,0) node[right] {$\bar{C}$};
        \draw[->] (0,-1.2) -- (0,4.2) node[left] {$\bar{p}$};

        \draw plot (\x, 0.2 + 0.5*\x*\x) node[right] {$\hatp =
          \frac{\barc^{2} T\mu |\mathbf{G}^{p}|}{|\mathbf{G}|} +
          \bars$};
        \draw[ultra thick,->] (1,2) node[above]
        {$(\barc^{\prime},\barpp)$} -- (1,1.5); \fill (1,2) circle
        [radius=0.4ex]; \draw[ultra thick,->] (1,2) -- (1.75,2);

        \draw[ultra thick,->] (3,1) node[below]
        {$(\barc^{\prime\prime},\barppp)$} -- (3,1.75); \fill (3,1)
        circle [radius=0.4ex]; \draw[ultra thick,->] (3,1) -- (2.5,1);
        
      \end{tikzpicture}
      \caption{The equilibrium preference as a function of $\barc$ for
        constant $\bark$.  Points on either side of this nullcline
        show component vectors of the solution of
        Equation~(\ref{eq:gbeta}).  }
      \label{fig:csquared}
    \end{center}
  
\end{figure}
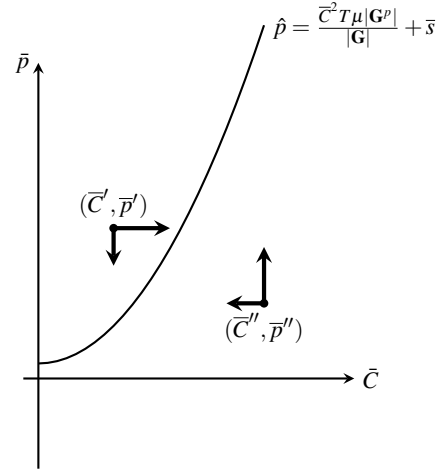

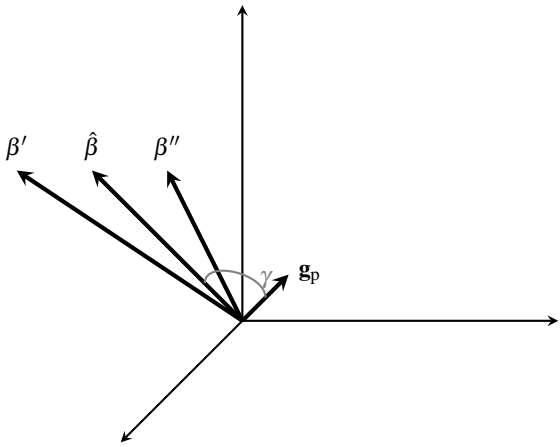
\begin{figure}[!]
    \begin{center}
      \begin{tikzpicture}[thick,>=stealth]
        \draw[->] (0,0,0) -- (4.2,0,0); \draw[->] (0,0,0) --
        (0,4.2,0); \draw[->] (0,0,0) -- (0,0,4.2);

        \draw[ultra thick,->] (0,0,0) -- (1,1,1) node[right]
        {$\gvec$};
        \draw[ultra thick,->] (0,0,0) -- (-3,2,0) node[above]
        {$\betavecp$}; \draw[ultra thick,->] (0,0,0) -- (-2,2,0)
        node[above] {$\betavechat$}; \draw[ultra thick,->] (0,0,0) --
        (-1,2,0) node[above] {$\betavecpp$};
        \draw[color=gray] (0.5,0.5,0.5) node[above] {$\gamma$}
        to[out=90,in=90] (-0.5,0.5,0);
      \end{tikzpicture}
      \caption{The position of three vectors describing
        Equation~(\ref{eq:app-dpdt}) at the points in
        Figure~\ref{fig:csquared} with a corresponding number of
        primes.  When $\dpdt = 0$ the two vectors are perpendicular,
        corresponding to the vector $\betavechat$.  When $\betavec$
        moves to $\betavecp$, $\gamma$ increases to greater than
        $\piovertwo$, meaning $\dpdt < 0$.  A similar argument shows
        that $\dpdt > 0$ when $\betavec$ moves to $\betavecpp$.}
      \label{fig:betavec}
    \end{center}
\end{figure}

I intend to show that the equilibria given in Equations~(\ref{eq:eq})
are asymptotically stable.  Particularly I wish to show that all
equilibria with $\barp > 0$ are asymptotically stable from any initial
condition if $\detGp > 0$.  Consider Figure~\ref{fig:csquared} where
$\hatp$ represents a quadratic function of $\hatc$ for some fixed
$\kappa$.  Choose a point $(\barc^{\prime},\barpp)$ above the curve
where $\barpp > \hatp$.  By representing the time derivative of
$\barp$ as a scalar product
\begin{equation}
  \label{eq:app-dpdt}
  \dpdt = \left(\frac{1}{2}\right)\left(\betavec \cdot \gvec\right) =
  \left(\frac{1}{2}\right)\left (|\betavec| |\gvec| \cos \gamma\right)
\end{equation}
where
\begin{equation}
  \gvec =
  \begin{pmatrix}
    G_{\kappa p} \\
    G_{p} \\
    G_{Cp}
  \end{pmatrix}
\end{equation}
we can more easily calculate the direction of the path in
Figure~\ref{fig:csquared}.  Equation~(\ref{eq:app-dpdt}) will be less
than zero if and only if $\gamma > \piovertwo$ (see
Figure~\ref{fig:betavec}).  At equilibrium the two vectors are
perpendicular and $\dpdt = 0$.  However, if we raise $\barp$ to
$\barpp$ then $\gamma > \piovertwo$ by making $\beta_{p}$ more
strongly negative (see Equation~(\ref{eq:beta-female})).  A similar
argument shows that $\dcdt > 0$ such that the solution will approach
the equilibrium depicted by the curve.  The only difference in the
argument lies in the non-zero equilibrium value of the scalar product
with $\mathbf{g_{\mathrm{C}}}$, defined analogously to $\gvec$.

Now consider the other side of the curve.  We can apply a similar
argument to show that $\dpdt > 0$ on this side of the curve.  Again at
equilibrium, $\gamma = \piovertwo$ and if we decrease $\barp$ to
$\barppp$ (see Figure~\ref{fig:csquared}) then we are increasing the
value of $\beta_{p}$ to greater than $\frac{T\mu \detGp}{\detG}$, such
that $\betavec$ is closer in $\mathbb{R}^{3}$ to $\gvec$.  Thus
$\gamma < \piovertwo$ and $\dpdt > 0$.  A similar argument in $\barc$
shows that $\dcdt < 0$ to the right of the curve.  The solution
approaches the curve in Figure~\ref{fig:csquared} regardless of the
magnitude of $T$ and $\mu$.

\end{document}

%% file: r_kp_01.tex
\begin{tikzpicture}[gnuplot]
\path (0.000,0.000) rectangle (7.620,5.080);
\gpcolor{color=gp lt color border}
\node[gp node right] at (1.136,0.985) { 0};
\node[gp node right] at (1.136,1.650) { 5};
\node[gp node right] at (1.136,2.316) { 10};
\node[gp node right] at (1.136,2.981) { 15};
\node[gp node right] at (1.136,3.646) { 20};
\node[gp node right] at (1.136,4.312) { 25};
\node[gp node center] at (1.320,0.677) {-0.2};
\node[gp node center] at (2.757,0.677) {-0.1};
\node[gp node center] at (4.194,0.677) { 0};
\node[gp node center] at (5.630,0.677) { 0.1};
\node[gp node center] at (7.067,0.677) { 0.2};
\gpsetlinetype{gp lt border}
\gpsetlinewidth{1.00}
\draw[gp path] (1.320,4.711)--(1.320,0.985)--(7.067,0.985)--(7.067,4.711)--cycle;
\node[gp node center,rotate=-270] at (0.246,2.848) {Size};
\node[gp node center] at (4.193,0.215) {Growth-preference Correlation ($\rho_{\kappa p}$)};
\gpcolor{rgb color={0.000,0.000,0.000}}
\gpsetlinetype{gp lt plot 0}
\gpsetlinewidth{4.00}
\draw[gp path] (1.320,2.633)--(1.378,2.591)--(1.436,2.550)--(1.494,2.509)--(1.552,2.470)%
  --(1.610,2.431)--(1.668,2.394)--(1.726,2.357)--(1.784,2.321)--(1.842,2.285)--(1.901,2.251)%
  --(1.959,2.217)--(2.017,2.184)--(2.075,2.152)--(2.133,2.121)--(2.191,2.090)--(2.249,2.060)%
  --(2.307,2.030)--(2.365,2.002)--(2.423,1.974)--(2.481,1.946)--(2.539,1.920)--(2.597,1.894)%
  --(2.655,1.868)--(2.713,1.843)--(2.771,1.819)--(2.829,1.795)--(2.887,1.772)--(2.945,1.750)%
  --(3.003,1.728)--(3.062,1.706)--(3.120,1.685)--(3.178,1.665)--(3.236,1.645)--(3.294,1.626)%
  --(3.352,1.607)--(3.410,1.588)--(3.468,1.570)--(3.526,1.553)--(3.584,1.536)--(3.642,1.519)%
  --(3.700,1.503)--(3.758,1.487)--(3.816,1.472)--(3.874,1.457)--(3.932,1.442)--(3.990,1.428)%
  --(4.048,1.415)--(4.106,1.401)--(4.164,1.388)--(4.223,1.376)--(4.281,1.363)--(4.339,1.351)%
  --(4.397,1.340)--(4.455,1.328)--(4.513,1.317)--(4.571,1.307)--(4.629,1.296)--(4.687,1.286)%
  --(4.745,1.277)--(4.803,1.267)--(4.861,1.258)--(4.919,1.249)--(4.977,1.241)--(5.035,1.232)%
  --(5.093,1.224)--(5.151,1.216)--(5.209,1.209)--(5.267,1.201)--(5.325,1.194)--(5.384,1.187)%
  --(5.442,1.181)--(5.500,1.174)--(5.558,1.168)--(5.616,1.162)--(5.674,1.156)--(5.732,1.151)%
  --(5.790,1.145)--(5.848,1.140)--(5.906,1.135)--(5.964,1.130)--(6.022,1.126)--(6.080,1.121)%
  --(6.138,1.117)--(6.196,1.113)--(6.254,1.109)--(6.312,1.105)--(6.370,1.101)--(6.428,1.098)%
  --(6.486,1.094)--(6.545,1.091)--(6.603,1.088)--(6.661,1.085)--(6.719,1.082)--(6.777,1.079)%
  --(6.835,1.077)--(6.893,1.074)--(6.951,1.072)--(7.009,1.070)--(7.067,1.068);
\gpcolor{color=gp lt color 1}
\gpsetlinetype{gp lt plot 1}
\draw[gp path] (1.320,3.639)--(1.378,3.586)--(1.436,3.535)--(1.494,3.484)--(1.552,3.435)%
  --(1.610,3.386)--(1.668,3.338)--(1.726,3.292)--(1.784,3.246)--(1.842,3.201)--(1.901,3.156)%
  --(1.959,3.113)--(2.017,3.071)--(2.075,3.029)--(2.133,2.988)--(2.191,2.948)--(2.249,2.908)%
  --(2.307,2.870)--(2.365,2.832)--(2.423,2.795)--(2.481,2.758)--(2.539,2.723)--(2.597,2.688)%
  --(2.655,2.653)--(2.713,2.620)--(2.771,2.587)--(2.829,2.554)--(2.887,2.523)--(2.945,2.492)%
  --(3.003,2.461)--(3.062,2.431)--(3.120,2.402)--(3.178,2.373)--(3.236,2.345)--(3.294,2.318)%
  --(3.352,2.291)--(3.410,2.264)--(3.468,2.238)--(3.526,2.213)--(3.584,2.188)--(3.642,2.164)%
  --(3.700,2.140)--(3.758,2.116)--(3.816,2.094)--(3.874,2.071)--(3.932,2.049)--(3.990,2.028)%
  --(4.048,2.006)--(4.106,1.986)--(4.164,1.966)--(4.223,1.946)--(4.281,1.926)--(4.339,1.907)%
  --(4.397,1.889)--(4.455,1.871)--(4.513,1.853)--(4.571,1.835)--(4.629,1.818)--(4.687,1.802)%
  --(4.745,1.785)--(4.803,1.769)--(4.861,1.753)--(4.919,1.738)--(4.977,1.723)--(5.035,1.708)%
  --(5.093,1.694)--(5.151,1.680)--(5.209,1.666)--(5.267,1.653)--(5.325,1.640)--(5.384,1.627)%
  --(5.442,1.614)--(5.500,1.602)--(5.558,1.590)--(5.616,1.578)--(5.674,1.566)--(5.732,1.555)%
  --(5.790,1.544)--(5.848,1.533)--(5.906,1.523)--(5.964,1.512)--(6.022,1.502)--(6.080,1.492)%
  --(6.138,1.483)--(6.196,1.473)--(6.254,1.464)--(6.312,1.455)--(6.370,1.446)--(6.428,1.438)%
  --(6.486,1.429)--(6.545,1.421)--(6.603,1.413)--(6.661,1.405)--(6.719,1.397)--(6.777,1.390)%
  --(6.835,1.383)--(6.893,1.375)--(6.951,1.368)--(7.009,1.362)--(7.067,1.355);
\gpcolor{color=gp lt color 2}
\gpsetlinetype{gp lt plot 2}
\draw[gp path] (1.320,1.251)--(1.378,1.251)--(1.436,1.251)--(1.494,1.251)--(1.552,1.251)%
  --(1.610,1.251)--(1.668,1.251)--(1.726,1.251)--(1.784,1.251)--(1.842,1.251)--(1.901,1.251)%
  --(1.959,1.251)--(2.017,1.251)--(2.075,1.251)--(2.133,1.251)--(2.191,1.251)--(2.249,1.251)%
  --(2.307,1.251)--(2.365,1.251)--(2.423,1.251)--(2.481,1.251)--(2.539,1.251)--(2.597,1.251)%
  --(2.655,1.251)--(2.713,1.251)--(2.771,1.251)--(2.829,1.251)--(2.887,1.251)--(2.945,1.251)%
  --(3.003,1.251)--(3.062,1.251)--(3.120,1.251)--(3.178,1.251)--(3.236,1.251)--(3.294,1.251)%
  --(3.352,1.251)--(3.410,1.251)--(3.468,1.251)--(3.526,1.251)--(3.584,1.251)--(3.642,1.251)%
  --(3.700,1.251)--(3.758,1.251)--(3.816,1.251)--(3.874,1.251)--(3.932,1.251)--(3.990,1.251)%
  --(4.048,1.251)--(4.106,1.251)--(4.164,1.251)--(4.223,1.251)--(4.281,1.251)--(4.339,1.251)%
  --(4.397,1.251)--(4.455,1.251)--(4.513,1.251)--(4.571,1.251)--(4.629,1.251)--(4.687,1.251)%
  --(4.745,1.251)--(4.803,1.251)--(4.861,1.251)--(4.919,1.251)--(4.977,1.251)--(5.035,1.251)%
  --(5.093,1.251)--(5.151,1.251)--(5.209,1.251)--(5.267,1.251)--(5.325,1.251)--(5.384,1.251)%
  --(5.442,1.251)--(5.500,1.251)--(5.558,1.251)--(5.616,1.251)--(5.674,1.251)--(5.732,1.251)%
  --(5.790,1.251)--(5.848,1.251)--(5.906,1.251)--(5.964,1.251)--(6.022,1.251)--(6.080,1.251)%
  --(6.138,1.251)--(6.196,1.251)--(6.254,1.251)--(6.312,1.251)--(6.370,1.251)--(6.428,1.251)%
  --(6.486,1.251)--(6.545,1.251)--(6.603,1.251)--(6.661,1.251)--(6.719,1.251)--(6.777,1.251)%
  --(6.835,1.251)--(6.893,1.251)--(6.951,1.251)--(7.009,1.251)--(7.067,1.251);
\gpcolor{color=gp lt color border}
\gpsetlinetype{gp lt border}
\gpsetlinewidth{1.00}
\draw[gp path] (1.320,4.711)--(1.320,0.985)--(7.067,0.985)--(7.067,4.711)--cycle;
\gpdefrectangularnode{gp plot 1}{\pgfpoint{1.320cm}{0.985cm}}{\pgfpoint{7.067cm}{4.711cm}}
\end{tikzpicture}

%% file: r_kp_02.tex
\begin{tikzpicture}[gnuplot]
\path (0.000,0.000) rectangle (7.620,5.080);
\gpcolor{color=gp lt color border}
\node[gp node right] at (1.136,0.985) { 0};
\node[gp node right] at (1.136,1.650) { 5};
\node[gp node right] at (1.136,2.316) { 10};
\node[gp node right] at (1.136,2.981) { 15};
\node[gp node right] at (1.136,3.646) { 20};
\node[gp node right] at (1.136,4.312) { 25};
\node[gp node center] at (1.320,0.677) {-0.2};
\node[gp node center] at (2.757,0.677) {-0.1};
\node[gp node center] at (4.194,0.677) { 0};
\node[gp node center] at (5.630,0.677) { 0.1};
\node[gp node center] at (7.067,0.677) { 0.2};
\gpsetlinetype{gp lt border}
\gpsetlinewidth{1.00}
\draw[gp path] (1.320,4.711)--(1.320,0.985)--(7.067,0.985)--(7.067,4.711)--cycle;
\node[gp node center,rotate=-270] at (0.246,2.848) {Size};
\node[gp node center] at (4.193,0.215) {Growth-preference Correlation ($\rho_{\kappa p}$)};
\gpcolor{rgb color={0.000,0.000,0.000}}
\gpsetlinetype{gp lt plot 0}
\gpsetlinewidth{4.00}
\draw[gp path] (1.320,1.616)--(1.378,1.621)--(1.436,1.626)--(1.494,1.631)--(1.552,1.636)%
  --(1.610,1.641)--(1.668,1.646)--(1.726,1.652)--(1.784,1.658)--(1.842,1.663)--(1.901,1.669)%
  --(1.959,1.675)--(2.017,1.682)--(2.075,1.688)--(2.133,1.694)--(2.191,1.701)--(2.249,1.708)%
  --(2.307,1.715)--(2.365,1.722)--(2.423,1.730)--(2.481,1.737)--(2.539,1.745)--(2.597,1.753)%
  --(2.655,1.761)--(2.713,1.769)--(2.771,1.778)--(2.829,1.786)--(2.887,1.795)--(2.945,1.804)%
  --(3.003,1.814)--(3.062,1.823)--(3.120,1.833)--(3.178,1.843)--(3.236,1.853)--(3.294,1.864)%
  --(3.352,1.875)--(3.410,1.886)--(3.468,1.897)--(3.526,1.908)--(3.584,1.920)--(3.642,1.932)%
  --(3.700,1.945)--(3.758,1.958)--(3.816,1.971)--(3.874,1.984)--(3.932,1.998)--(3.990,2.012)%
  --(4.048,2.026)--(4.106,2.041)--(4.164,2.055)--(4.223,2.071)--(4.281,2.087)--(4.339,2.103)%
  --(4.397,2.119)--(4.455,2.136)--(4.513,2.153)--(4.571,2.171)--(4.629,2.189)--(4.687,2.207)%
  --(4.745,2.226)--(4.803,2.245)--(4.861,2.265)--(4.919,2.285)--(4.977,2.306)--(5.035,2.327)%
  --(5.093,2.349)--(5.151,2.371)--(5.209,2.393)--(5.267,2.416)--(5.325,2.440)--(5.384,2.464)%
  --(5.442,2.489)--(5.500,2.514)--(5.558,2.540)--(5.616,2.566)--(5.674,2.593)--(5.732,2.621)%
  --(5.790,2.649)--(5.848,2.678)--(5.906,2.707)--(5.964,2.737)--(6.022,2.768)--(6.080,2.799)%
  --(6.138,2.831)--(6.196,2.864)--(6.254,2.897)--(6.312,2.932)--(6.370,2.966)--(6.428,3.002)%
  --(6.486,3.038)--(6.545,3.076)--(6.603,3.114)--(6.661,3.152)--(6.719,3.192)--(6.777,3.232)%
  --(6.835,3.273)--(6.893,3.316)--(6.951,3.359)--(7.009,3.402)--(7.067,3.447);
\gpcolor{color=gp lt color 1}
\gpsetlinetype{gp lt plot 1}
\draw[gp path] (1.320,1.903)--(1.378,1.913)--(1.436,1.922)--(1.494,1.932)--(1.552,1.942)%
  --(1.610,1.952)--(1.668,1.962)--(1.726,1.972)--(1.784,1.983)--(1.842,1.993)--(1.901,2.004)%
  --(1.959,2.015)--(2.017,2.027)--(2.075,2.038)--(2.133,2.050)--(2.191,2.062)--(2.249,2.074)%
  --(2.307,2.086)--(2.365,2.099)--(2.423,2.112)--(2.481,2.125)--(2.539,2.138)--(2.597,2.151)%
  --(2.655,2.165)--(2.713,2.179)--(2.771,2.193)--(2.829,2.208)--(2.887,2.223)--(2.945,2.238)%
  --(3.003,2.253)--(3.062,2.268)--(3.120,2.284)--(3.178,2.300)--(3.236,2.317)--(3.294,2.334)%
  --(3.352,2.351)--(3.410,2.368)--(3.468,2.386)--(3.526,2.404)--(3.584,2.422)--(3.642,2.441)%
  --(3.700,2.460)--(3.758,2.479)--(3.816,2.499)--(3.874,2.519)--(3.932,2.540)--(3.990,2.561)%
  --(4.048,2.582)--(4.106,2.604)--(4.164,2.626)--(4.223,2.648)--(4.281,2.671)--(4.339,2.695)%
  --(4.397,2.718)--(4.455,2.743)--(4.513,2.767)--(4.571,2.792)--(4.629,2.818)--(4.687,2.844)%
  --(4.745,2.871)--(4.803,2.898)--(4.861,2.925)--(4.919,2.953)--(4.977,2.982)--(5.035,3.011)%
  --(5.093,3.041)--(5.151,3.071)--(5.209,3.102)--(5.267,3.133)--(5.325,3.165)--(5.384,3.198)%
  --(5.442,3.231)--(5.500,3.264)--(5.558,3.299)--(5.616,3.334)--(5.674,3.370)--(5.732,3.406)%
  --(5.790,3.443)--(5.848,3.481)--(5.906,3.519)--(5.964,3.558)--(6.022,3.598)--(6.080,3.638)%
  --(6.138,3.680)--(6.196,3.722)--(6.254,3.765)--(6.312,3.808)--(6.370,3.853)--(6.428,3.898)%
  --(6.486,3.944)--(6.545,3.991)--(6.603,4.039)--(6.661,4.087)--(6.719,4.137)--(6.777,4.187)%
  --(6.835,4.238)--(6.893,4.291)--(6.951,4.344)--(7.009,4.398)--(7.067,4.453);
\gpcolor{color=gp lt color 2}
\gpsetlinetype{gp lt plot 2}
\draw[gp path] (1.320,1.251)--(1.378,1.251)--(1.436,1.251)--(1.494,1.251)--(1.552,1.251)%
  --(1.610,1.251)--(1.668,1.251)--(1.726,1.251)--(1.784,1.251)--(1.842,1.251)--(1.901,1.251)%
  --(1.959,1.251)--(2.017,1.251)--(2.075,1.251)--(2.133,1.251)--(2.191,1.251)--(2.249,1.251)%
  --(2.307,1.251)--(2.365,1.251)--(2.423,1.251)--(2.481,1.251)--(2.539,1.251)--(2.597,1.251)%
  --(2.655,1.251)--(2.713,1.251)--(2.771,1.251)--(2.829,1.251)--(2.887,1.251)--(2.945,1.251)%
  --(3.003,1.251)--(3.062,1.251)--(3.120,1.251)--(3.178,1.251)--(3.236,1.251)--(3.294,1.251)%
  --(3.352,1.251)--(3.410,1.251)--(3.468,1.251)--(3.526,1.251)--(3.584,1.251)--(3.642,1.251)%
  --(3.700,1.251)--(3.758,1.251)--(3.816,1.251)--(3.874,1.251)--(3.932,1.251)--(3.990,1.251)%
  --(4.048,1.251)--(4.106,1.251)--(4.164,1.251)--(4.223,1.251)--(4.281,1.251)--(4.339,1.251)%
  --(4.397,1.251)--(4.455,1.251)--(4.513,1.251)--(4.571,1.251)--(4.629,1.251)--(4.687,1.251)%
  --(4.745,1.251)--(4.803,1.251)--(4.861,1.251)--(4.919,1.251)--(4.977,1.251)--(5.035,1.251)%
  --(5.093,1.251)--(5.151,1.251)--(5.209,1.251)--(5.267,1.251)--(5.325,1.251)--(5.384,1.251)%
  --(5.442,1.251)--(5.500,1.251)--(5.558,1.251)--(5.616,1.251)--(5.674,1.251)--(5.732,1.251)%
  --(5.790,1.251)--(5.848,1.251)--(5.906,1.251)--(5.964,1.251)--(6.022,1.251)--(6.080,1.251)%
  --(6.138,1.251)--(6.196,1.251)--(6.254,1.251)--(6.312,1.251)--(6.370,1.251)--(6.428,1.251)%
  --(6.486,1.251)--(6.545,1.251)--(6.603,1.251)--(6.661,1.251)--(6.719,1.251)--(6.777,1.251)%
  --(6.835,1.251)--(6.893,1.251)--(6.951,1.251)--(7.009,1.251)--(7.067,1.251);
\gpcolor{color=gp lt color border}
\gpsetlinetype{gp lt border}
\gpsetlinewidth{1.00}
\draw[gp path] (1.320,4.711)--(1.320,0.985)--(7.067,0.985)--(7.067,4.711)--cycle;
\gpdefrectangularnode{gp plot 1}{\pgfpoint{1.320cm}{0.985cm}}{\pgfpoint{7.067cm}{4.711cm}}
\end{tikzpicture}

%% file: r_kp_03.tex
\begin{tikzpicture}[gnuplot]
\path (0.000,0.000) rectangle (7.620,5.080);
\gpcolor{color=gp lt color border}
\node[gp node right] at (0.952,0.985) { 0};
\node[gp node right] at (0.952,1.730) { 1};
\node[gp node right] at (0.952,2.475) { 2};
\node[gp node right] at (0.952,3.221) { 3};
\node[gp node right] at (0.952,3.966) { 4};
\node[gp node right] at (0.952,4.711) { 5};
\node[gp node center] at (1.136,0.677) {-0.2};
\node[gp node center] at (2.619,0.677) {-0.1};
\node[gp node center] at (4.102,0.677) { 0};
\node[gp node center] at (5.584,0.677) { 0.1};
\node[gp node center] at (7.067,0.677) { 0.2};
\gpsetlinetype{gp lt border}
\gpsetlinewidth{1.00}
\draw[gp path] (1.136,4.711)--(1.136,0.985)--(7.067,0.985)--(7.067,4.711)--cycle;
\node[gp node center,rotate=-270] at (0.246,2.848) {Size};
\node[gp node center] at (4.101,0.215) {Growth-preference Correlation ($\rho_{\kappa p}$)};
\gpcolor{rgb color={0.000,0.000,0.000}}
\gpsetlinetype{gp lt plot 0}
\gpsetlinewidth{4.00}
\draw[gp path] (1.136,1.885)--(1.196,1.884)--(1.256,1.884)--(1.316,1.884)--(1.376,1.883)%
  --(1.436,1.883)--(1.495,1.883)--(1.555,1.883)--(1.615,1.882)--(1.675,1.882)--(1.735,1.882)%
  --(1.795,1.882)--(1.855,1.882)--(1.915,1.882)--(1.975,1.882)--(2.035,1.882)--(2.095,1.882)%
  --(2.154,1.882)--(2.214,1.883)--(2.274,1.883)--(2.334,1.883)--(2.394,1.883)--(2.454,1.884)%
  --(2.514,1.884)--(2.574,1.884)--(2.634,1.885)--(2.694,1.885)--(2.754,1.885)--(2.813,1.886)%
  --(2.873,1.886)--(2.933,1.887)--(2.993,1.887)--(3.053,1.888)--(3.113,1.888)--(3.173,1.889)%
  --(3.233,1.889)--(3.293,1.890)--(3.353,1.891)--(3.413,1.891)--(3.472,1.892)--(3.532,1.893)%
  --(3.592,1.893)--(3.652,1.894)--(3.712,1.895)--(3.772,1.896)--(3.832,1.897)--(3.892,1.897)%
  --(3.952,1.898)--(4.012,1.899)--(4.072,1.900)--(4.131,1.901)--(4.191,1.902)--(4.251,1.903)%
  --(4.311,1.904)--(4.371,1.904)--(4.431,1.905)--(4.491,1.906)--(4.551,1.907)--(4.611,1.908)%
  --(4.671,1.909)--(4.731,1.910)--(4.790,1.911)--(4.850,1.912)--(4.910,1.914)--(4.970,1.915)%
  --(5.030,1.916)--(5.090,1.917)--(5.150,1.918)--(5.210,1.919)--(5.270,1.920)--(5.330,1.921)%
  --(5.390,1.922)--(5.449,1.924)--(5.509,1.925)--(5.569,1.926)--(5.629,1.927)--(5.689,1.928)%
  --(5.749,1.930)--(5.809,1.931)--(5.869,1.932)--(5.929,1.933)--(5.989,1.934)--(6.049,1.936)%
  --(6.108,1.937)--(6.168,1.938)--(6.228,1.939)--(6.288,1.941)--(6.348,1.942)--(6.408,1.943)%
  --(6.468,1.945)--(6.528,1.946)--(6.588,1.947)--(6.648,1.948)--(6.708,1.950)--(6.767,1.951)%
  --(6.827,1.952)--(6.887,1.954)--(6.947,1.955)--(7.007,1.956)--(7.067,1.958);
\gpcolor{color=gp lt color 1}
\gpsetlinetype{gp lt plot 1}
\draw[gp path] (1.136,3.012)--(1.196,3.000)--(1.256,2.988)--(1.316,2.976)--(1.376,2.964)%
  --(1.436,2.953)--(1.495,2.941)--(1.555,2.930)--(1.615,2.919)--(1.675,2.907)--(1.735,2.896)%
  --(1.795,2.886)--(1.855,2.875)--(1.915,2.864)--(1.975,2.854)--(2.035,2.843)--(2.095,2.833)%
  --(2.154,2.822)--(2.214,2.812)--(2.274,2.802)--(2.334,2.792)--(2.394,2.783)--(2.454,2.773)%
  --(2.514,2.763)--(2.574,2.754)--(2.634,2.744)--(2.694,2.735)--(2.754,2.726)--(2.813,2.717)%
  --(2.873,2.708)--(2.933,2.699)--(2.993,2.690)--(3.053,2.681)--(3.113,2.673)--(3.173,2.664)%
  --(3.233,2.656)--(3.293,2.647)--(3.353,2.639)--(3.413,2.631)--(3.472,2.623)--(3.532,2.615)%
  --(3.592,2.607)--(3.652,2.599)--(3.712,2.591)--(3.772,2.584)--(3.832,2.576)--(3.892,2.569)%
  --(3.952,2.561)--(4.012,2.554)--(4.072,2.547)--(4.131,2.540)--(4.191,2.532)--(4.251,2.526)%
  --(4.311,2.519)--(4.371,2.512)--(4.431,2.505)--(4.491,2.498)--(4.551,2.492)--(4.611,2.485)%
  --(4.671,2.479)--(4.731,2.473)--(4.790,2.466)--(4.850,2.460)--(4.910,2.454)--(4.970,2.448)%
  --(5.030,2.442)--(5.090,2.436)--(5.150,2.430)--(5.210,2.425)--(5.270,2.419)--(5.330,2.413)%
  --(5.390,2.408)--(5.449,2.402)--(5.509,2.397)--(5.569,2.392)--(5.629,2.386)--(5.689,2.381)%
  --(5.749,2.376)--(5.809,2.371)--(5.869,2.366)--(5.929,2.361)--(5.989,2.356)--(6.049,2.351)%
  --(6.108,2.347)--(6.168,2.342)--(6.228,2.337)--(6.288,2.333)--(6.348,2.328)--(6.408,2.324)%
  --(6.468,2.320)--(6.528,2.315)--(6.588,2.311)--(6.648,2.307)--(6.708,2.303)--(6.767,2.299)%
  --(6.827,2.295)--(6.887,2.291)--(6.947,2.287)--(7.007,2.283)--(7.067,2.280);
\gpcolor{color=gp lt color 2}
\gpsetlinetype{gp lt plot 2}
\draw[gp path] (1.136,2.475)--(1.196,2.475)--(1.256,2.475)--(1.316,2.475)--(1.376,2.475)%
  --(1.436,2.475)--(1.495,2.475)--(1.555,2.475)--(1.615,2.475)--(1.675,2.475)--(1.735,2.475)%
  --(1.795,2.475)--(1.855,2.475)--(1.915,2.475)--(1.975,2.475)--(2.035,2.475)--(2.095,2.475)%
  --(2.154,2.475)--(2.214,2.475)--(2.274,2.475)--(2.334,2.475)--(2.394,2.475)--(2.454,2.475)%
  --(2.514,2.475)--(2.574,2.475)--(2.634,2.475)--(2.694,2.475)--(2.754,2.475)--(2.813,2.475)%
  --(2.873,2.475)--(2.933,2.475)--(2.993,2.475)--(3.053,2.475)--(3.113,2.475)--(3.173,2.475)%
  --(3.233,2.475)--(3.293,2.475)--(3.353,2.475)--(3.413,2.475)--(3.472,2.475)--(3.532,2.475)%
  --(3.592,2.475)--(3.652,2.475)--(3.712,2.475)--(3.772,2.475)--(3.832,2.475)--(3.892,2.475)%
  --(3.952,2.475)--(4.012,2.475)--(4.072,2.475)--(4.131,2.475)--(4.191,2.475)--(4.251,2.475)%
  --(4.311,2.475)--(4.371,2.475)--(4.431,2.475)--(4.491,2.475)--(4.551,2.475)--(4.611,2.475)%
  --(4.671,2.475)--(4.731,2.475)--(4.790,2.475)--(4.850,2.475)--(4.910,2.475)--(4.970,2.475)%
  --(5.030,2.475)--(5.090,2.475)--(5.150,2.475)--(5.210,2.475)--(5.270,2.475)--(5.330,2.475)%
  --(5.390,2.475)--(5.449,2.475)--(5.509,2.475)--(5.569,2.475)--(5.629,2.475)--(5.689,2.475)%
  --(5.749,2.475)--(5.809,2.475)--(5.869,2.475)--(5.929,2.475)--(5.989,2.475)--(6.049,2.475)%
  --(6.108,2.475)--(6.168,2.475)--(6.228,2.475)--(6.288,2.475)--(6.348,2.475)--(6.408,2.475)%
  --(6.468,2.475)--(6.528,2.475)--(6.588,2.475)--(6.648,2.475)--(6.708,2.475)--(6.767,2.475)%
  --(6.827,2.475)--(6.887,2.475)--(6.947,2.475)--(7.007,2.475)--(7.067,2.475);
\gpcolor{color=gp lt color border}
\gpsetlinetype{gp lt border}
\gpsetlinewidth{1.00}
\draw[gp path] (1.136,4.711)--(1.136,0.985)--(7.067,0.985)--(7.067,4.711)--cycle;
\gpdefrectangularnode{gp plot 1}{\pgfpoint{1.136cm}{0.985cm}}{\pgfpoint{7.067cm}{4.711cm}}
\end{tikzpicture}

%% file: r_kp_04.tex
\begin{tikzpicture}[gnuplot]
\path (0.000,0.000) rectangle (7.620,5.080);
\gpcolor{color=gp lt color border}
\node[gp node right] at (0.952,0.985) { 0};
\node[gp node right] at (0.952,1.730) { 1};
\node[gp node right] at (0.952,2.475) { 2};
\node[gp node right] at (0.952,3.221) { 3};
\node[gp node right] at (0.952,3.966) { 4};
\node[gp node right] at (0.952,4.711) { 5};
\node[gp node center] at (1.136,0.677) {-0.2};
\node[gp node center] at (2.619,0.677) {-0.1};
\node[gp node center] at (4.102,0.677) { 0};
\node[gp node center] at (5.584,0.677) { 0.1};
\node[gp node center] at (7.067,0.677) { 0.2};
\gpsetlinetype{gp lt border}
\gpsetlinewidth{1.00}
\draw[gp path] (1.136,4.711)--(1.136,0.985)--(7.067,0.985)--(7.067,4.711)--cycle;
\node[gp node center,rotate=-270] at (0.246,2.848) {Size};
\node[gp node center] at (4.101,0.215) {Growth-preference Correlation ($\rho_{\kappa p}$)};
\gpcolor{rgb color={0.000,0.000,0.000}}
\gpsetlinetype{gp lt plot 0}
\gpsetlinewidth{4.00}
\draw[gp path] (1.136,3.001)--(1.196,3.003)--(1.256,3.004)--(1.316,3.006)--(1.376,3.008)%
  --(1.436,3.009)--(1.495,3.011)--(1.555,3.013)--(1.615,3.015)--(1.675,3.016)--(1.735,3.018)%
  --(1.795,3.020)--(1.855,3.021)--(1.915,3.023)--(1.975,3.025)--(2.035,3.027)--(2.095,3.028)%
  --(2.154,3.030)--(2.214,3.032)--(2.274,3.034)--(2.334,3.035)--(2.394,3.037)--(2.454,3.039)%
  --(2.514,3.041)--(2.574,3.043)--(2.634,3.044)--(2.694,3.046)--(2.754,3.048)--(2.813,3.050)%
  --(2.873,3.052)--(2.933,3.054)--(2.993,3.056)--(3.053,3.058)--(3.113,3.059)--(3.173,3.061)%
  --(3.233,3.063)--(3.293,3.065)--(3.353,3.067)--(3.413,3.069)--(3.472,3.071)--(3.532,3.073)%
  --(3.592,3.075)--(3.652,3.077)--(3.712,3.079)--(3.772,3.081)--(3.832,3.083)--(3.892,3.085)%
  --(3.952,3.087)--(4.012,3.089)--(4.072,3.092)--(4.131,3.094)--(4.191,3.096)--(4.251,3.098)%
  --(4.311,3.100)--(4.371,3.102)--(4.431,3.105)--(4.491,3.107)--(4.551,3.109)--(4.611,3.111)%
  --(4.671,3.114)--(4.731,3.116)--(4.790,3.118)--(4.850,3.121)--(4.910,3.123)--(4.970,3.125)%
  --(5.030,3.128)--(5.090,3.130)--(5.150,3.133)--(5.210,3.135)--(5.270,3.138)--(5.330,3.140)%
  --(5.390,3.143)--(5.449,3.145)--(5.509,3.148)--(5.569,3.151)--(5.629,3.153)--(5.689,3.156)%
  --(5.749,3.159)--(5.809,3.161)--(5.869,3.164)--(5.929,3.167)--(5.989,3.170)--(6.049,3.173)%
  --(6.108,3.176)--(6.168,3.178)--(6.228,3.181)--(6.288,3.184)--(6.348,3.187)--(6.408,3.190)%
  --(6.468,3.193)--(6.528,3.197)--(6.588,3.200)--(6.648,3.203)--(6.708,3.206)--(6.767,3.209)%
  --(6.827,3.213)--(6.887,3.216)--(6.947,3.219)--(7.007,3.223)--(7.067,3.226);
\gpcolor{color=gp lt color 1}
\gpsetlinetype{gp lt plot 1}
\draw[gp path] (1.136,3.323)--(1.196,3.330)--(1.256,3.337)--(1.316,3.343)--(1.376,3.350)%
  --(1.436,3.357)--(1.495,3.364)--(1.555,3.371)--(1.615,3.379)--(1.675,3.386)--(1.735,3.393)%
  --(1.795,3.400)--(1.855,3.408)--(1.915,3.415)--(1.975,3.423)--(2.035,3.431)--(2.095,3.438)%
  --(2.154,3.446)--(2.214,3.454)--(2.274,3.462)--(2.334,3.469)--(2.394,3.477)--(2.454,3.486)%
  --(2.514,3.494)--(2.574,3.502)--(2.634,3.510)--(2.694,3.518)--(2.754,3.527)--(2.813,3.535)%
  --(2.873,3.544)--(2.933,3.552)--(2.993,3.561)--(3.053,3.570)--(3.113,3.579)--(3.173,3.588)%
  --(3.233,3.597)--(3.293,3.606)--(3.353,3.615)--(3.413,3.624)--(3.472,3.633)--(3.532,3.643)%
  --(3.592,3.652)--(3.652,3.662)--(3.712,3.671)--(3.772,3.681)--(3.832,3.691)--(3.892,3.700)%
  --(3.952,3.710)--(4.012,3.720)--(4.072,3.730)--(4.131,3.740)--(4.191,3.751)--(4.251,3.761)%
  --(4.311,3.771)--(4.371,3.782)--(4.431,3.792)--(4.491,3.803)--(4.551,3.814)--(4.611,3.825)%
  --(4.671,3.836)--(4.731,3.847)--(4.790,3.858)--(4.850,3.869)--(4.910,3.880)--(4.970,3.892)%
  --(5.030,3.903)--(5.090,3.915)--(5.150,3.926)--(5.210,3.938)--(5.270,3.950)--(5.330,3.962)%
  --(5.390,3.974)--(5.449,3.986)--(5.509,3.998)--(5.569,4.010)--(5.629,4.023)--(5.689,4.035)%
  --(5.749,4.048)--(5.809,4.061)--(5.869,4.074)--(5.929,4.086)--(5.989,4.100)--(6.049,4.113)%
  --(6.108,4.126)--(6.168,4.139)--(6.228,4.153)--(6.288,4.166)--(6.348,4.180)--(6.408,4.194)%
  --(6.468,4.208)--(6.528,4.222)--(6.588,4.236)--(6.648,4.250)--(6.708,4.264)--(6.767,4.279)%
  --(6.827,4.293)--(6.887,4.308)--(6.947,4.323)--(7.007,4.338)--(7.067,4.353);
\gpcolor{color=gp lt color 2}
\gpsetlinetype{gp lt plot 2}
\draw[gp path] (1.136,2.475)--(1.196,2.475)--(1.256,2.475)--(1.316,2.475)--(1.376,2.475)%
  --(1.436,2.475)--(1.495,2.475)--(1.555,2.475)--(1.615,2.475)--(1.675,2.475)--(1.735,2.475)%
  --(1.795,2.475)--(1.855,2.475)--(1.915,2.475)--(1.975,2.475)--(2.035,2.475)--(2.095,2.475)%
  --(2.154,2.475)--(2.214,2.475)--(2.274,2.475)--(2.334,2.475)--(2.394,2.475)--(2.454,2.475)%
  --(2.514,2.475)--(2.574,2.475)--(2.634,2.475)--(2.694,2.475)--(2.754,2.475)--(2.813,2.475)%
  --(2.873,2.475)--(2.933,2.475)--(2.993,2.475)--(3.053,2.475)--(3.113,2.475)--(3.173,2.475)%
  --(3.233,2.475)--(3.293,2.475)--(3.353,2.475)--(3.413,2.475)--(3.472,2.475)--(3.532,2.475)%
  --(3.592,2.475)--(3.652,2.475)--(3.712,2.475)--(3.772,2.475)--(3.832,2.475)--(3.892,2.475)%
  --(3.952,2.475)--(4.012,2.475)--(4.072,2.475)--(4.131,2.475)--(4.191,2.475)--(4.251,2.475)%
  --(4.311,2.475)--(4.371,2.475)--(4.431,2.475)--(4.491,2.475)--(4.551,2.475)--(4.611,2.475)%
  --(4.671,2.475)--(4.731,2.475)--(4.790,2.475)--(4.850,2.475)--(4.910,2.475)--(4.970,2.475)%
  --(5.030,2.475)--(5.090,2.475)--(5.150,2.475)--(5.210,2.475)--(5.270,2.475)--(5.330,2.475)%
  --(5.390,2.475)--(5.449,2.475)--(5.509,2.475)--(5.569,2.475)--(5.629,2.475)--(5.689,2.475)%
  --(5.749,2.475)--(5.809,2.475)--(5.869,2.475)--(5.929,2.475)--(5.989,2.475)--(6.049,2.475)%
  --(6.108,2.475)--(6.168,2.475)--(6.228,2.475)--(6.288,2.475)--(6.348,2.475)--(6.408,2.475)%
  --(6.468,2.475)--(6.528,2.475)--(6.588,2.475)--(6.648,2.475)--(6.708,2.475)--(6.767,2.475)%
  --(6.827,2.475)--(6.887,2.475)--(6.947,2.475)--(7.007,2.475)--(7.067,2.475);
\gpcolor{color=gp lt color border}
\gpsetlinetype{gp lt border}
\gpsetlinewidth{1.00}
\draw[gp path] (1.136,4.711)--(1.136,0.985)--(7.067,0.985)--(7.067,4.711)--cycle;
\gpdefrectangularnode{gp plot 1}{\pgfpoint{1.136cm}{0.985cm}}{\pgfpoint{7.067cm}{4.711cm}}
\end{tikzpicture}

%% file: tpeqb01_pos.tex
\begin{tikzpicture}[gnuplot]
\path (0.000,0.000) rectangle (12.500,8.750);
\gpcolor{color=gp lt color border}
\node[gp node right] at (1.136,0.985) { 0};
\node[gp node right] at (1.136,2.464) { 2};
\node[gp node right] at (1.136,3.943) { 4};
\node[gp node right] at (1.136,5.423) { 6};
\node[gp node right] at (1.136,6.902) { 8};
\node[gp node right] at (1.136,8.381) { 10};
\node[gp node center] at (1.879,0.677) { 2};
\node[gp node center] at (2.998,0.677) { 4};
\node[gp node center] at (4.117,0.677) { 6};
\node[gp node center] at (5.235,0.677) { 8};
\node[gp node center] at (6.354,0.677) { 10};
\node[gp node center] at (7.472,0.677) { 12};
\node[gp node center] at (8.591,0.677) { 14};
\node[gp node center] at (9.710,0.677) { 16};
\node[gp node center] at (10.828,0.677) { 18};
\node[gp node center] at (11.947,0.677) { 20};
\gpsetlinetype{gp lt border}
\gpsetlinewidth{1.00}
\draw[gp path] (1.320,8.381)--(1.320,0.985)--(11.947,0.985)--(11.947,8.381)--cycle;
\node[gp node center,rotate=-270] at (0.246,4.683) {Size};
\node[gp node center] at (6.633,0.215) {Generation Time ($T$)};
\gpcolor{rgb color={0.000,0.000,0.000}}
\gpsetlinetype{gp lt plot 0}
\gpsetlinewidth{4.00}
\draw[gp path] (2.313,8.381)--(2.393,7.651)--(2.501,6.897)--(2.608,6.304)--(2.715,5.830)%
  --(2.823,5.447)--(2.930,5.133)--(3.037,4.874)--(3.145,4.658)--(3.252,4.476)--(3.360,4.322)%
  --(3.467,4.191)--(3.574,4.078)--(3.682,3.981)--(3.789,3.897)--(3.896,3.824)--(4.004,3.760)%
  --(4.111,3.704)--(4.218,3.654)--(4.326,3.611)--(4.433,3.573)--(4.540,3.539)--(4.648,3.509)%
  --(4.755,3.482)--(4.862,3.459)--(4.970,3.438)--(5.077,3.419)--(5.184,3.403)--(5.292,3.388)%
  --(5.399,3.375)--(5.506,3.364)--(5.614,3.354)--(5.721,3.345)--(5.828,3.337)--(5.936,3.331)%
  --(6.043,3.325)--(6.150,3.320)--(6.258,3.316)--(6.365,3.313)--(6.472,3.310)--(6.580,3.308)%
  --(6.687,3.306)--(6.795,3.305)--(6.902,3.305)--(7.009,3.305)--(7.117,3.305)--(7.224,3.306)%
  --(7.331,3.307)--(7.439,3.308)--(7.546,3.310)--(7.653,3.312)--(7.761,3.314)--(7.868,3.317)%
  --(7.975,3.320)--(8.083,3.323)--(8.190,3.326)--(8.297,3.329)--(8.405,3.333)--(8.512,3.337)%
  --(8.619,3.341)--(8.727,3.345)--(8.834,3.349)--(8.941,3.354)--(9.049,3.358)--(9.156,3.363)%
  --(9.263,3.368)--(9.371,3.373)--(9.478,3.378)--(9.585,3.383)--(9.693,3.388)--(9.800,3.394)%
  --(9.907,3.399)--(10.015,3.405)--(10.122,3.410)--(10.230,3.416)--(10.337,3.422)--(10.444,3.428)%
  --(10.552,3.434)--(10.659,3.440)--(10.766,3.446)--(10.874,3.452)--(10.981,3.458)--(11.088,3.465)%
  --(11.196,3.471)--(11.303,3.477)--(11.410,3.484)--(11.518,3.490)--(11.625,3.497)--(11.732,3.503)%
  --(11.840,3.510)--(11.947,3.517);
\gpcolor{color=gp lt color 1}
\gpsetlinetype{gp lt plot 1}
\draw[gp path] (3.075,8.381)--(3.145,8.134)--(3.252,7.802)--(3.360,7.511)--(3.467,7.253)%
  --(3.574,7.024)--(3.682,6.818)--(3.789,6.634)--(3.896,6.467)--(4.004,6.316)--(4.111,6.178)%
  --(4.218,6.052)--(4.326,5.936)--(4.433,5.830)--(4.540,5.732)--(4.648,5.641)--(4.755,5.557)%
  --(4.862,5.479)--(4.970,5.407)--(5.077,5.339)--(5.184,5.276)--(5.292,5.217)--(5.399,5.162)%
  --(5.506,5.110)--(5.614,5.061)--(5.721,5.016)--(5.828,4.973)--(5.936,4.932)--(6.043,4.894)%
  --(6.150,4.858)--(6.258,4.824)--(6.365,4.791)--(6.472,4.761)--(6.580,4.732)--(6.687,4.705)%
  --(6.795,4.679)--(6.902,4.654)--(7.009,4.631)--(7.117,4.609)--(7.224,4.588)--(7.331,4.568)%
  --(7.439,4.549)--(7.546,4.531)--(7.653,4.514)--(7.761,4.498)--(7.868,4.483)--(7.975,4.468)%
  --(8.083,4.455)--(8.190,4.442)--(8.297,4.429)--(8.405,4.417)--(8.512,4.406)--(8.619,4.395)%
  --(8.727,4.385)--(8.834,4.376)--(8.941,4.367)--(9.049,4.358)--(9.156,4.350)--(9.263,4.343)%
  --(9.371,4.335)--(9.478,4.329)--(9.585,4.322)--(9.693,4.316)--(9.800,4.311)--(9.907,4.305)%
  --(10.015,4.300)--(10.122,4.296)--(10.230,4.292)--(10.337,4.288)--(10.444,4.284)--(10.552,4.280)%
  --(10.659,4.277)--(10.766,4.274)--(10.874,4.272)--(10.981,4.269)--(11.088,4.267)--(11.196,4.265)%
  --(11.303,4.264)--(11.410,4.262)--(11.518,4.261)--(11.625,4.260)--(11.732,4.259)--(11.840,4.258)%
  --(11.947,4.258);
\gpcolor{color=gp lt color 2}
\gpsetlinetype{gp lt plot 2}
\draw[gp path] (1.320,2.464)--(1.427,2.464)--(1.535,2.464)--(1.642,2.464)--(1.749,2.464)%
  --(1.857,2.464)--(1.964,2.464)--(2.071,2.464)--(2.179,2.464)--(2.286,2.464)--(2.393,2.464)%
  --(2.501,2.464)--(2.608,2.464)--(2.715,2.464)--(2.823,2.464)--(2.930,2.464)--(3.037,2.464)%
  --(3.145,2.464)--(3.252,2.464)--(3.360,2.464)--(3.467,2.464)--(3.574,2.464)--(3.682,2.464)%
  --(3.789,2.464)--(3.896,2.464)--(4.004,2.464)--(4.111,2.464)--(4.218,2.464)--(4.326,2.464)%
  --(4.433,2.464)--(4.540,2.464)--(4.648,2.464)--(4.755,2.464)--(4.862,2.464)--(4.970,2.464)%
  --(5.077,2.464)--(5.184,2.464)--(5.292,2.464)--(5.399,2.464)--(5.506,2.464)--(5.614,2.464)%
  --(5.721,2.464)--(5.828,2.464)--(5.936,2.464)--(6.043,2.464)--(6.150,2.464)--(6.258,2.464)%
  --(6.365,2.464)--(6.472,2.464)--(6.580,2.464)--(6.687,2.464)--(6.795,2.464)--(6.902,2.464)%
  --(7.009,2.464)--(7.117,2.464)--(7.224,2.464)--(7.331,2.464)--(7.439,2.464)--(7.546,2.464)%
  --(7.653,2.464)--(7.761,2.464)--(7.868,2.464)--(7.975,2.464)--(8.083,2.464)--(8.190,2.464)%
  --(8.297,2.464)--(8.405,2.464)--(8.512,2.464)--(8.619,2.464)--(8.727,2.464)--(8.834,2.464)%
  --(8.941,2.464)--(9.049,2.464)--(9.156,2.464)--(9.263,2.464)--(9.371,2.464)--(9.478,2.464)%
  --(9.585,2.464)--(9.693,2.464)--(9.800,2.464)--(9.907,2.464)--(10.015,2.464)--(10.122,2.464)%
  --(10.230,2.464)--(10.337,2.464)--(10.444,2.464)--(10.552,2.464)--(10.659,2.464)--(10.766,2.464)%
  --(10.874,2.464)--(10.981,2.464)--(11.088,2.464)--(11.196,2.464)--(11.303,2.464)--(11.410,2.464)%
  --(11.518,2.464)--(11.625,2.464)--(11.732,2.464)--(11.840,2.464)--(11.947,2.464);
\gpcolor{color=gp lt color border}
\gpsetlinetype{gp lt border}
\gpsetlinewidth{1.00}
\draw[gp path] (1.320,8.381)--(1.320,0.985)--(11.947,0.985)--(11.947,8.381)--cycle;
\gpdefrectangularnode{gp plot 1}{\pgfpoint{1.320cm}{0.985cm}}{\pgfpoint{11.947cm}{8.381cm}}
\end{tikzpicture}

%% file: tpeqb01_neg.tex
\begin{tikzpicture}[gnuplot]
\path (0.000,0.000) rectangle (12.500,8.750);
\gpcolor{color=gp lt color border}
\node[gp node right] at (1.136,0.985) { 0};
\node[gp node right] at (1.136,2.464) { 2};
\node[gp node right] at (1.136,3.943) { 4};
\node[gp node right] at (1.136,5.423) { 6};
\node[gp node right] at (1.136,6.902) { 8};
\node[gp node right] at (1.136,8.381) { 10};
\node[gp node center] at (1.879,0.677) { 2};
\node[gp node center] at (2.998,0.677) { 4};
\node[gp node center] at (4.117,0.677) { 6};
\node[gp node center] at (5.235,0.677) { 8};
\node[gp node center] at (6.354,0.677) { 10};
\node[gp node center] at (7.472,0.677) { 12};
\node[gp node center] at (8.591,0.677) { 14};
\node[gp node center] at (9.710,0.677) { 16};
\node[gp node center] at (10.828,0.677) { 18};
\node[gp node center] at (11.947,0.677) { 20};
\gpsetlinetype{gp lt border}
\gpsetlinewidth{1.00}
\draw[gp path] (1.320,8.381)--(1.320,0.985)--(11.947,0.985)--(11.947,8.381)--cycle;
\node[gp node center,rotate=-270] at (0.246,4.683) {Size};
\node[gp node center] at (6.633,0.215) {Generation Time ($T$)};
\gpcolor{rgb color={0.000,0.000,0.000}}
\gpsetlinetype{gp lt plot 0}
\gpsetlinewidth{4.00}
\draw[gp path] (1.320,6.358)--(1.427,5.548)--(1.535,5.032)--(1.642,4.674)--(1.749,4.413)%
  --(1.857,4.214)--(1.964,4.058)--(2.071,3.932)--(2.179,3.829)--(2.286,3.743)--(2.393,3.671)%
  --(2.501,3.609)--(2.608,3.557)--(2.715,3.511)--(2.823,3.471)--(2.930,3.437)--(3.037,3.406)%
  --(3.145,3.379)--(3.252,3.356)--(3.360,3.335)--(3.467,3.317)--(3.574,3.300)--(3.682,3.286)%
  --(3.789,3.273)--(3.896,3.262)--(4.004,3.252)--(4.111,3.243)--(4.218,3.235)--(4.326,3.229)%
  --(4.433,3.223)--(4.540,3.218)--(4.648,3.214)--(4.755,3.211)--(4.862,3.208)--(4.970,3.205)%
  --(5.077,3.204)--(5.184,3.203)--(5.292,3.202)--(5.399,3.201)--(5.506,3.202)--(5.614,3.202)%
  --(5.721,3.203)--(5.828,3.204)--(5.936,3.206)--(6.043,3.207)--(6.150,3.209)--(6.258,3.212)%
  --(6.365,3.214)--(6.472,3.217)--(6.580,3.220)--(6.687,3.223)--(6.795,3.226)--(6.902,3.230)%
  --(7.009,3.234)--(7.117,3.237)--(7.224,3.242)--(7.331,3.246)--(7.439,3.250)--(7.546,3.254)%
  --(7.653,3.259)--(7.761,3.264)--(7.868,3.269)--(7.975,3.273)--(8.083,3.279)--(8.190,3.284)%
  --(8.297,3.289)--(8.405,3.294)--(8.512,3.300)--(8.619,3.305)--(8.727,3.311)--(8.834,3.316)%
  --(8.941,3.322)--(9.049,3.328)--(9.156,3.334)--(9.263,3.340)--(9.371,3.346)--(9.478,3.352)%
  --(9.585,3.358)--(9.693,3.364)--(9.800,3.370)--(9.907,3.377)--(10.015,3.383)--(10.122,3.389)%
  --(10.230,3.396)--(10.337,3.402)--(10.444,3.409)--(10.552,3.415)--(10.659,3.422)--(10.766,3.429)%
  --(10.874,3.435)--(10.981,3.442)--(11.088,3.449)--(11.196,3.456)--(11.303,3.463)--(11.410,3.469)%
  --(11.518,3.476)--(11.625,3.483)--(11.732,3.490)--(11.840,3.497)--(11.947,3.504);
\gpcolor{color=gp lt color 1}
\gpsetlinetype{gp lt plot 1}
\draw[gp path] (1.398,8.381)--(1.427,8.035)--(1.535,7.173)--(1.642,6.555)--(1.749,6.090)%
  --(1.857,5.727)--(1.964,5.435)--(2.071,5.197)--(2.179,4.998)--(2.286,4.830)--(2.393,4.686)%
  --(2.501,4.562)--(2.608,4.454)--(2.715,4.359)--(2.823,4.275)--(2.930,4.201)--(3.037,4.134)%
  --(3.145,4.075)--(3.252,4.021)--(3.360,3.973)--(3.467,3.929)--(3.574,3.889)--(3.682,3.853)%
  --(3.789,3.820)--(3.896,3.790)--(4.004,3.763)--(4.111,3.738)--(4.218,3.715)--(4.326,3.694)%
  --(4.433,3.674)--(4.540,3.657)--(4.648,3.640)--(4.755,3.626)--(4.862,3.612)--(4.970,3.599)%
  --(5.077,3.588)--(5.184,3.577)--(5.292,3.568)--(5.399,3.559)--(5.506,3.551)--(5.614,3.544)%
  --(5.721,3.537)--(5.828,3.531)--(5.936,3.526)--(6.043,3.521)--(6.150,3.517)--(6.258,3.513)%
  --(6.365,3.510)--(6.472,3.507)--(6.580,3.505)--(6.687,3.503)--(6.795,3.501)--(6.902,3.500)%
  --(7.009,3.499)--(7.117,3.498)--(7.224,3.498)--(7.331,3.498)--(7.439,3.498)--(7.546,3.499)%
  --(7.653,3.500)--(7.761,3.501)--(7.868,3.502)--(7.975,3.503)--(8.083,3.505)--(8.190,3.507)%
  --(8.297,3.509)--(8.405,3.511)--(8.512,3.513)--(8.619,3.516)--(8.727,3.519)--(8.834,3.522)%
  --(8.941,3.525)--(9.049,3.528)--(9.156,3.531)--(9.263,3.535)--(9.371,3.538)--(9.478,3.542)%
  --(9.585,3.546)--(9.693,3.550)--(9.800,3.554)--(9.907,3.558)--(10.015,3.562)--(10.122,3.566)%
  --(10.230,3.571)--(10.337,3.575)--(10.444,3.580)--(10.552,3.585)--(10.659,3.590)--(10.766,3.594)%
  --(10.874,3.599)--(10.981,3.604)--(11.088,3.609)--(11.196,3.615)--(11.303,3.620)--(11.410,3.625)%
  --(11.518,3.630)--(11.625,3.636)--(11.732,3.641)--(11.840,3.647)--(11.947,3.653);
\gpcolor{color=gp lt color 2}
\gpsetlinetype{gp lt plot 2}
\draw[gp path] (1.320,2.464)--(1.427,2.464)--(1.535,2.464)--(1.642,2.464)--(1.749,2.464)%
  --(1.857,2.464)--(1.964,2.464)--(2.071,2.464)--(2.179,2.464)--(2.286,2.464)--(2.393,2.464)%
  --(2.501,2.464)--(2.608,2.464)--(2.715,2.464)--(2.823,2.464)--(2.930,2.464)--(3.037,2.464)%
  --(3.145,2.464)--(3.252,2.464)--(3.360,2.464)--(3.467,2.464)--(3.574,2.464)--(3.682,2.464)%
  --(3.789,2.464)--(3.896,2.464)--(4.004,2.464)--(4.111,2.464)--(4.218,2.464)--(4.326,2.464)%
  --(4.433,2.464)--(4.540,2.464)--(4.648,2.464)--(4.755,2.464)--(4.862,2.464)--(4.970,2.464)%
  --(5.077,2.464)--(5.184,2.464)--(5.292,2.464)--(5.399,2.464)--(5.506,2.464)--(5.614,2.464)%
  --(5.721,2.464)--(5.828,2.464)--(5.936,2.464)--(6.043,2.464)--(6.150,2.464)--(6.258,2.464)%
  --(6.365,2.464)--(6.472,2.464)--(6.580,2.464)--(6.687,2.464)--(6.795,2.464)--(6.902,2.464)%
  --(7.009,2.464)--(7.117,2.464)--(7.224,2.464)--(7.331,2.464)--(7.439,2.464)--(7.546,2.464)%
  --(7.653,2.464)--(7.761,2.464)--(7.868,2.464)--(7.975,2.464)--(8.083,2.464)--(8.190,2.464)%
  --(8.297,2.464)--(8.405,2.464)--(8.512,2.464)--(8.619,2.464)--(8.727,2.464)--(8.834,2.464)%
  --(8.941,2.464)--(9.049,2.464)--(9.156,2.464)--(9.263,2.464)--(9.371,2.464)--(9.478,2.464)%
  --(9.585,2.464)--(9.693,2.464)--(9.800,2.464)--(9.907,2.464)--(10.015,2.464)--(10.122,2.464)%
  --(10.230,2.464)--(10.337,2.464)--(10.444,2.464)--(10.552,2.464)--(10.659,2.464)--(10.766,2.464)%
  --(10.874,2.464)--(10.981,2.464)--(11.088,2.464)--(11.196,2.464)--(11.303,2.464)--(11.410,2.464)%
  --(11.518,2.464)--(11.625,2.464)--(11.732,2.464)--(11.840,2.464)--(11.947,2.464);
\gpcolor{color=gp lt color border}
\gpsetlinetype{gp lt border}
\gpsetlinewidth{1.00}
\draw[gp path] (1.320,8.381)--(1.320,0.985)--(11.947,0.985)--(11.947,8.381)--cycle;
\gpdefrectangularnode{gp plot 1}{\pgfpoint{1.320cm}{0.985cm}}{\pgfpoint{11.947cm}{8.381cm}}
\end{tikzpicture}